\documentclass[twocolumn, usenatbib]{mnras}
\usepackage{savesym}
\usepackage{graphicx}
\usepackage{longtable}
\usepackage{changepage}

\expandafter\let\csname equation*\endcsname\relax
  \expandafter\let\csname endequation*\endcsname\relax 
\usepackage{subfig}
\usepackage{amsmath}
\usepackage{amssymb}
\usepackage{verbatim}
\usepackage[yyyymmdd,hhmmss]{datetime}
\usepackage{array}
\usepackage{times}
\usepackage{xcolor}

\DeclareRobustCommand{\DE}[3]{#2}
\let\DEthebibliography\thebibliography
\def\thebibliography{\DeclareRobustCommand{\DE}[3]{##3}\DEthebibliography}

\title[Accretion disc variability at high energies]{ The turbulent variability of accretion discs observed at high energies }
\author [Andrew Mummery \& Samuel G. D. Turner]{Andrew Mummery$^1$\thanks{E-mail:
andrew.mummery@physics.ox.ac.uk} and Samuel G. D. Turner$^{2, 3}$
\\
$^1$ Oxford Theoretical Physics, Beecroft Building,  Clarendon Laboratory, Parks Road, Oxford, OX1 3PU, UK \\ 
$^2$ Department of Applied Mathematics and Theoretical Physics, University of Cambridge, Wilberforce Road, Cambridge, CB3 0WA, UK \\
$^3$ Institute of Astronomy, University of Cambridge, Madingley Road, Cambridge, CB3 0HA, UK }

\date{}

\begin{document}

\pagerange{\pageref{firstpage}--\pageref{lastpage}} \pubyear{2024}

\maketitle

\label{firstpage}

\begin{abstract} 
We use numerical stochastic-viscous hydrodynamic simulations and new analytical results from thin disc theory to probe the turbulent variability of accretion flows, as observed at high energies. We show that the act of observing accretion discs in the Wien tail exponentially enhances small-scale temperature variability in the flow, which in a real disc will be driven by magnetohydrodynamic turbulence, to large amplitude luminosity fluctuations (as predicted analytically). In particular, we demonstrate that discs with more spatially coherent turbulence (as might be expected of thicker discs), and relativistic discs observed at larger inclinations, show significantly enhancement in their Wien-tail variability. 
We believe this is the first analysis of relativistic viewing-angle effects on {turbulent} variability in the literature. Using these results we argue that tidal disruption events represent particularly interesting systems with which to study accretion flow variability, and may in fact be the best astrophysical probes of small scale disc turbulence. This is a result of a typical tidal disruption event disc being naturally observed in the Wien-tail and likely having a somewhat thicker disc and cleaner X-ray spectrum than other sources. We argue for dedicated X-ray observational campaigns of tidal disruption events, with the aim of studying accretion flow variability.  
\end{abstract}

\noindent

\section{Introduction} 
Accretion, whereby the energy of infalling material is liberated by turbulent dissipation and ultimately radiated away, is one of the most ubiquitous and important phenomena in the universe. The central object powering the accretion process can range from young stellar objects (YSOs), white dwarfs (WDs) to black holes (BHs). Similarly, there are many different possible mechanisms for the origin of the accreted material, including Roche-lobe overflow in cataclysmic variables (CVs) and X-ray binaries (XRBs), tidal disruption events (TDEs) where a star on a close orbit of a BH is tidally disrupted, and the accretion of galactic gas by supermassive BHs (SMBHs) that power active galactic nuclei (AGN). Understanding the nature of the accretion process is important not only for understanding these objects, but also because of the {role} accretion has in wider astrophysics, for example in the triggering of Type Ia supernovae and in the effect feedback from AGN has on the evolution of their host galaxies.

One interesting property of these systems is that, despite the large range of scales that they cover, they show remarkably similar variability properties. These include a log-normal luminosity distribution; a linear relation between the mean flux varying over long timescales and the root mean square (rms) variability on short timescales; broad noise across a range of frequencies as seen in the power spectral density (PSD); and hard and soft lags observed between different energy bands. The presence of these features has been observed in YSOs \citep{Scaringi2015}, CVs \citep{Scaringi2012a,  VeresvarskaScaringi2023}, XRBs \citep{Nowak2000, Gleissner2004, Gandhi2009} and AGN \citep{LyutyiOknyanskii1987, Gaskell2004, Arevalo2006, Markowitz2007, Fabian2009, DeMarco2011, Vaughan2011}. 

The theory of propagating fluctuations, first proposed by \cite{Lyubarskii97}, is often used to explain these observed phenomena. In this theory, local fluctuations in the accretion rate are created at all radii. These fluctuations then propagate inwards\footnote{Some fraction of the mass fluctuation must also propagate outwards to conserve angular momentum, although this is often neglected in modelling.}, combining with those from smaller radii. Assuming that the timescale these fluctuations are created on increases at larger radii in the disc (note that, in a Newtonian disc, all the key timescales scale as ${r^{3/2}}$ where $r$ is the radial distance from the central object), this naturally creates a broad PSD across a range of frequencies. Further assuming that the fluctuations combine multiplicatively, this also gives rise to a linear rms-flux relation, as the fractional variation created by the high frequency fluctuations in the inner disc provide a constant fractional modulation to the low frequency fluctuations from the outer disc. This linear rms-flux relation, provided that it holds for all frequencies, implies that the underlying lightcurve must be log-normally distributed \citep{Uttley05}. As fluctuations propagate inwards, they will first pass through the cooler, outer regions of the disc, before the hotter, inner regions. This will naturally create hard lags, as any signal will appear first in the lower energy bands. \citet{Mushtukov2018} additionally showed that, if the theory is extended to allow for any fluctuations to propagate outwards, then this can also lead to soft lags. All of these results also hold in a relativistic theory of propagating fluctuations \citep{Mummery23b}. 

Since the seminal work of \cite{Lyubarskii97}, there has been extensive analytic work on propagating fluctuations \citep[e.g.][]{Kotov2001, ArevaloUttley2006, IngramDone2011, IngramVanDerKlis2013, Mushtukov2018, Mummery23b}. Additionally, numerical studies have allowed the model to be tested in the non-linear regime. These have been performed in 1D (based on the standard disc diffusion equation with a stochastic $\alpha$ parameter) \citep{CowperthwaiteReynolds2014, Turner21}, and have also recently been generalised to 2D \citep{Turner23}. Additionally, high resolution magnetohydronamic (MHD) simulations have been found to exhibit behaviour which is consistent with the propagating fluctuations picture \citep{Hogg16, Bollimpalli2020}.  

One of the more recent theoretical disc variability predictions (not precisely reliant on the theory of propagating fluctuations, merely the log-normality of disc temperature variability) was derived by \cite{MummeryBalbus22}, who highlighted a key result from thin disc theory. \cite{MummeryBalbus22} argued that when observed in the Wien tail (i.e., at photon energies larger than the peak temperature of the disc), fluctuations in the disc temperature are exponentially amplified into significantly larger luminosity fluctuations. This is formally relevant for all disc systems, but in particular for discs formed in tidal disruption events, which typically peak at temperature $kT_p \lesssim 100$ eV, a factor of a few below typical X-ray bands (with lower bandpass energies typically $E_l = 300$ eV). 
As these systems are observed in their Wien-tail, and therefore may well have exponentially enhanced variability,  these could be the ideal disc systems to observe and probe our theoretical descriptions of disc variability in an astronomical setting. 

Before comparing to observations however, it is imperative to ascertain a more precise set of predictions of the modern theory of propagating fluctuations \citep[in particular the numerical simulations of][]{Turner23}, when filtered through the Wien-tail modelling of \cite{MummeryBalbus22}. Deriving observational predictions is the purpose of this paper. We shall do this by taking the numerical models of \cite{Turner23}, specialising to regions of parameter space relevant for observations of tidal disruption events. We then ``observe'' these systems, demonstrating that the analytic theory of \cite{MummeryBalbus22} reproduces the numerical results well, allowing us to calibrate observational predictions.  

We will demonstrate that, if the canonical theory of propagating fluctuations is correct, that tidal disruption events should be observed to show large, and potentially up to an order of magnitude, fluctuations of their X-ray luminosity on timescales as short as days--weeks (or even potentially hours for lower mass black holes). This means that tidal disruption events represent ideal probes of the small-scale disc turbulence in black hole accretion discs. 

The layout of the paper is as follows. In section \ref{num_mod} we recap the work of \cite{Turner23}, and describe the numerical two-dimensional viscous-hydrodynamic model used throughout this paper. In section \ref{analy_sec} we recap the analytical model of \cite{MummeryBalbus22}, and show that the analytical model accurately reproduces the full numerical simulations. We then demonstrate that different assumptions regarding the properties of the stochastic viscosity (in particular the length scale over which the viscosity is coherent) produce pronounced effects on the Wien-tail lightcurves, and in particular we argue that thicker discs will produce significantly larger variability. In section \ref{rel_sec} we highlight relativistic viewing angle effects, and show that {\it at fixed observed X-ray luminosity} more inclined relativistic disc systems will be observed to be more variable than face on systems (at the factor 2 level). In section \ref{discussion} we discuss the relevance of our results for observations of tidal disruption events, before concluding in section \ref{conclusion}. 

\section{ Numerical model }\label{num_mod}

The numerical model analysed in this work is presented in detail in \cite{Turner23}, where a two-dimensional (vertically integrated) generalisation of the one-dimensional model presented in \cite{Turner21} was derived and analysed. We only briefly summarise the fundamental equations solved by the model here, and direct the interested reader to these papers for further information and discussion. 

The disc fluid is assumed to evolve under the equations of viscous hydrodynamics, with a Newtonian gravitational potential. Explicitly, the governing equations are mass conservation (written in Cartesian index notation, where repeated indices are summed over)
\begin{equation}\label{mass_conc}
    {\partial \rho \over \partial t} +  {\partial \over \partial x_j} (\rho  v_j) = 0, 
\end{equation}
and momentum conservation 
\begin{equation}
    {\partial \over \partial t} (\rho v_i) + {\partial \over \partial x_j} (\rho  v_i  v_j + p \, \delta_{ij} ) = - \rho  {\partial \over \partial x_i}  \Phi +  {\partial \over \partial x_j}  \Pi_{ij} . 
\end{equation}
In these expressions $\rho$ is the disc density, $\vec v$ the disc velocity, $p$ the disc pressure and $\delta_{ij}$ is the Kronecker delta. We use the notation $x_i$ for the Cartesian coordinate system and $t$ is time as usual.  The viscous stress tensor is given by 
\begin{equation}
    \Pi_{ij} = \rho \nu \left[ {\partial v_j \over \partial x_i} +  {\partial v_i \over \partial x_j}- {2\over 3}  {\partial v_k \over \partial x_k}  \, \delta_{ij} \right] ,
\end{equation}
and we assume a standard central point source gravitational potential 
\begin{equation}\label{gravity}
    \Phi = - {GM \over R} ,
\end{equation}
where $G$ is the Gravitational constant, $M$ the mass of the central BH and $R$ the spherical radius. As is now standard, the coefficient of disc viscosity is set by the \cite{SS73} $\alpha$-prescription, namely 
\begin{equation}
    \nu = \alpha c_s H, 
\end{equation}
where $c_s = \sqrt{p/\rho}$ and $H$ are the disc's speed of sound and pressure scale height respectively.

The work of \citet{Turner23} further assumed that the disc is sufficiently thin (i.e. ${H\ll r}$, where $r$ is the cylindrical radius in the plane of the disc) that the exact details of the vertical structure can be ignored, allowing the problem to be confined to a 2D, vertically integrated approach. This allows the density $\rho$ to be replaced with the surface density $\Sigma$. Formally, this is done (in cylindrical polar coordinates $(r,\phi,z)$ which each have their standard meaning) as
\begin{equation}
    \Sigma(r,\phi) = \int^\infty_{-\infty} \rho(r,\phi,z){\rm d}z\, ,
\end{equation}
but this can be thought of simply as $\Sigma\equiv \rho H$. A similar equation defines the vertically integrated pressure, ${P\equiv pH}$. It is implicitly assumed that other variables, such as the fluid velocity, are not a function of $z$. The thin disc assumption also allows $R$ to be replaced with $r$ in eq. \ref{gravity}. 

In order to close the equations, the (approximate) solution of vertical hydrostatic equilibrium is used to relate the sound speed to the disc scale height as
\begin{equation}
    c_s^2 = {GMH^2 \over r^3}\, .
\end{equation}
The disc aspect ratio ${\cal H} \equiv H/r$ is taken as an input parameter of the model, which also sets the sound speed. Note that the sound speed varies as $r^{-1/2}$ through the disc. The simulations analysed in this work have an aspect ratio of ${\cal H} = 0.1$, although other aspect ratios were explored by \citet{Turner23}.

Unlike traditional approaches, $\alpha$ is not taken to be constant in this analysis. Instead, in an attempt to mimic the effects of the real magnetohydrodynamic turbulence in the disc, the $\alpha$-parameter is taken to be a stochastic random variable, and is equal to 
\begin{equation}\label{alpha_beta}
    \alpha(r, \phi, t) = \alpha_0 \exp\left[\beta(r, \phi, t)\right] , 
\end{equation}
where a value of $\alpha_0=0.1$ is used in all the simulations of \citet{Turner23}. $\beta$ is a (two-dimensional) stochastic random variable, which is evolved stochastically in time and is additionally advected with the flow, in an effort to mimic the properties of the magneto-rotational instability \citep{BalbusHawley91}, and the subsequent turbulent evolution of the disc. To be precise,  $\beta$ is evolved according to an Ornstein-Uhlenbeck (hereafter OU) process. An OU process can be thought of as a generalisation of a random walk to continuous time, in which in addition the properties of the process have been chosen so that there is a tendency for the parameter to move back towards a mean value, with a growing attraction to the mean as the distance between the parameter and its mean increases. 

At a given point $(r, \phi)$ in the disc, the instantaneous stochastic evolution under an OU process is given by
\begin{equation}\label{beta_OU}
    {\rm d} \beta = -\omega (\beta - \mu) {\rm d}t + \xi {\rm d}W ,
\end{equation}
where $1/\omega$ is the characteristic timescale of the OU process, $\mu$ is the mean value of $\beta$, ${\rm d}W \sim N (0, {\rm d}t)$ is a random normal variable and $\xi$ is a constant which determines the magnitude of the variation, which can be written in terms of the rms-amplitude of $\beta$ as ${\xi = \sqrt{2\omega \left\langle \beta^2 \right\rangle}}$. All the simulations presented in \citet{Turner23} took $\mu = 0$ and $\left\langle \beta^2 \right\rangle = 1$\footnote{Other values of $\left\langle \beta^2 \right\rangle$ were tested in \citet{Turner23} but the only effect was found to be on the normalisation of the variability and so no specific results were presented from these other simulations. }. Note that the choice of $\mu=0$ does not mean that $\langle\alpha\rangle=\alpha_0$ since the exponential function in eq. \ref{alpha_beta} has a positive skew.

Multiple driving timescales/frequencies were tested in \citet{Turner23} (and an even wider range tested in the 1D work of \citet{Turner21}), but the simulations used in this work use the fiducial driving frequency of 
\begin{equation}\label{timescale}
    \omega = \alpha_0 \left({GM\over r^3}\right)^{1/2} .
\end{equation}
This corresponds to a timescale which is 10 ($1/\alpha_0$) times longer than the orbital timescale, and was chosen to be similar to the MRI dynamo timescale which appears to be the most important timescale for the evolution of the effective $\alpha$ parameter in MHD simulations \citep[e.g.][]{Hogg16}.

In addition to the required normality of the random variable ${\rm d}W$, the exact form was chosen such that this random noise was spatially coherent over a length scale $l_c$ within the disc. Motivated by the nature of a mean-field theory ``turbulent viscosity'', we might expect that (to first order) turbulent eddies within the disc will have a size given approximately by the disc thickness. Therefore, the fiducial assumption used in \citet{Turner23} is that $l_c=H$, although other simulations where $l_c$ is some multiple of $H$ were performed and are used in this work. For the full details of the nature of the ${\rm d}W$ and the general stochastic prescription, we directed the interested reader to Appendix A within \citet{Turner23}.

In addition to being a stochastic variable, the parameter $\beta$ is assumed to be advected with the flow. In other words, the existing $\beta$ field is evolved with time through 
\begin{equation}\label{beta_adv}
{\partial \beta \over \partial t} + v_j {\partial \beta \over \partial x_j}  = 0    .
\end{equation}
Under eq. \ref{beta_adv}, existing structures within the $\beta$ field are evolved, and regions of high (or low) $\beta$ will be sheared out due to the presence of differential rotation within the disc. Eqs. \ref{beta_OU} and \ref{beta_adv} completely describe the evolution of $\beta$, and ultimately the local viscous stress tensor $\Pi_{ij}$ of the flow. 

The coupled equations (\ref{mass_conc}--\ref{beta_adv}) are solved by the code {\tt PLUTO} \cite{Mignone07}. Note that while these equations have been presented in Cartesian notation, they are solved in polar coordinates $(r,\phi)$. The initial condition of the disc surface density is that of the analytical one-dimensional steady state solution \cite{Pringle81} 
\begin{equation} \label{steady_state}
    \Sigma(r, \phi, 0) = {\dot M_0 \over 3\pi \nu} \left(1 - \sqrt{r_\star \over r}\right),
\end{equation}
where $r_\star = 6GM/c^2$ is the inner edge of the disc, set by the innermost stable circular orbit (ISCO) of a Schwarzschild BH.  Additionally, the velocity field was initialised with $v_\phi$ equal to the local Keplerian velocity $v_\phi = \sqrt{GM/r}$ with $v_r = 0$ and $\beta = 0$ everywhere. The fluid is first evolved during a run-in phase in which the stochasticity is turned on (equations \ref{beta_OU} and \ref{beta_adv}). This phase lasts for a total length of time $t= 0.4 \times 10^6 t_g$, where $t_g \equiv GM/c^3$.  This ensures that the $\beta$ field will have settled into a statistically steady state before we start taking results. The disc then continues to evolve forward for a further $t = 1.6\times 10^6 t_g$, where results are taken. We save the full state of the inner disc, defined by $6 < rc^2/GM < 1000$,  each $\Delta t = 1000 t_g$. We therefore have 1600 snapshots of the disc for which to analyse. 

\subsection{ Emergent disc spectrum }
\begin{figure}
    \centering
    \includegraphics[width=\linewidth]{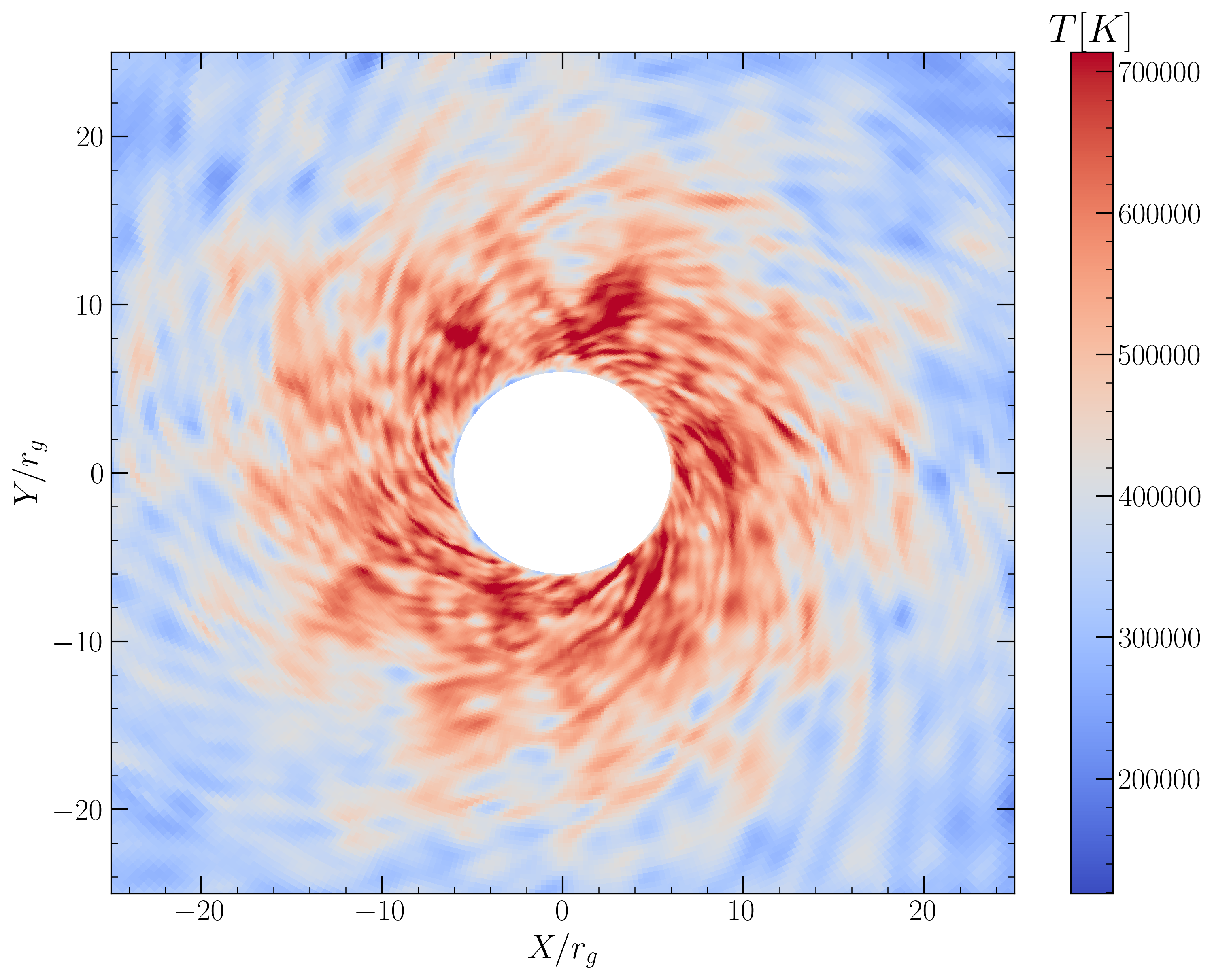}
    \caption{ An example snapshot of the two dimensional ($x$ and $y$) disc temperature profile. The central mass was taken to be $M = 2 \times 10^6 M_\odot$, and the average mass accretion rate $\dot M_0 = 6 \times 10^{20}$ kg/s. It is clear that the disc temperature is a two dimensional stochastic random field. There is small length scale coherence to the temperature field, which in this case is of order the disc scale height $l_c \sim H = 0.1 r$. The spectrum and lightcurves which result from observing this stochastic temperature field inherit the temperature's underlying randomness. }
    \label{fig:example_snapshot}
\end{figure}

In cylindrical coordinates with no $z$ dependence, the local energy liberated from viscous dissipation  $D(r, \phi, t)$ is equal to 
\begin{equation}
    \label{eq:diss}
    \begin{aligned}
        D(r,\phi,t) = \frac{1}{2}\nu&\Sigma\Bigg\{
        2\left[\left({\partial v_r \over \partial r}\right)^2 + \left(\frac{1}{r}{\partial v_\phi \over \partial \phi} + \frac{v_r}{r}\right)^2\right] \\
        & \left[ r{\partial \over \partial r}\left(\frac{v_\phi}{r}\right) + \frac{1}{r}{\partial v_r \over \partial \phi} \right]^2
        - \frac{2}{3}\left(\vec\nabla\cdot\vec {v}\right)^2\Bigg\}\, ,
    \end{aligned}
\end{equation}
where the factor of $1/2$ comes from the two (upper and lower) surfaces of the disc. Assuming that this dissipated energy is radiated locally, this defines the usual effective temperature $T(r, \phi, t)$ of the disc by 
\begin{equation}
    \sigma T^4 \equiv D(r,\phi,t), 
\end{equation}
where $\sigma$ is the usual Stefan-Boltzmann constant. An example snapshot of temperature of the inner disc is shown in Fig. \ref{fig:example_snapshot}. The temperature in this Figure is re-scaled to a system corresponding to a black hole of mass $M = 2 \times 10^6 M_\odot$, and average accretion rate $\dot M_0 = 6\times 10^{20}$ kg/s \citep[{\tt PLUTO} solves the equations in a dimensionless unit system, which can then always be rescaled by appropriate choice of $M$ and $\dot M_0$;][]{Turner23}. 

Clearly it can be seen in Fig. \ref{fig:example_snapshot} that the disc temperature is a stochastic random variable, with small length scale coherence to the temperature field (in this simulation the coherence length is equal to the disc scale height $l_c = H = 0.1 r$). While the average (over azimuth) temperature profile approximately follows what would be expected from one-dimensional thin disc theory, the local structure of the disc is much more complicated.  We shall go on to demonstrate that at low observing energies this two-dimensional structure effectively averages out, but that this is not true at high observing frequencies (compared to the peak disc temperature).

In a purely Newtonian theory of gravity (i.e., one without gravitational or Doppler energy shifting of photons, which are assumed to travel on straight lines), the emergent spectrum (energy received per unit time per unit area per unit photon energy) can be written 
\begin{equation}\label{newt_spec}
    F_E(E, t) = {\cos i\over D^2} \int_{r_\star}^\infty \int_0^{2\pi} {2 E^3 \over h^3 c^2} { r \, {\rm d}\phi \, {\rm d} r  \over \exp(E/kT) - 1} , 
\end{equation}
where $h$ and $k$ are the Planck and Boltzmann constants respectively,  $D$ is the distance to the observer, and $i$ is the inclination of the disc as observed by the distant observer. The spectral luminosity is defined by $L_E \equiv 4\pi D^2 F_E$, and is independent of the observers distance. For each snapshot of the disc temperature $T(r, \phi, t)$, the above integral can be performed. 

\begin{figure}
    \centering
    \includegraphics[width=\linewidth]{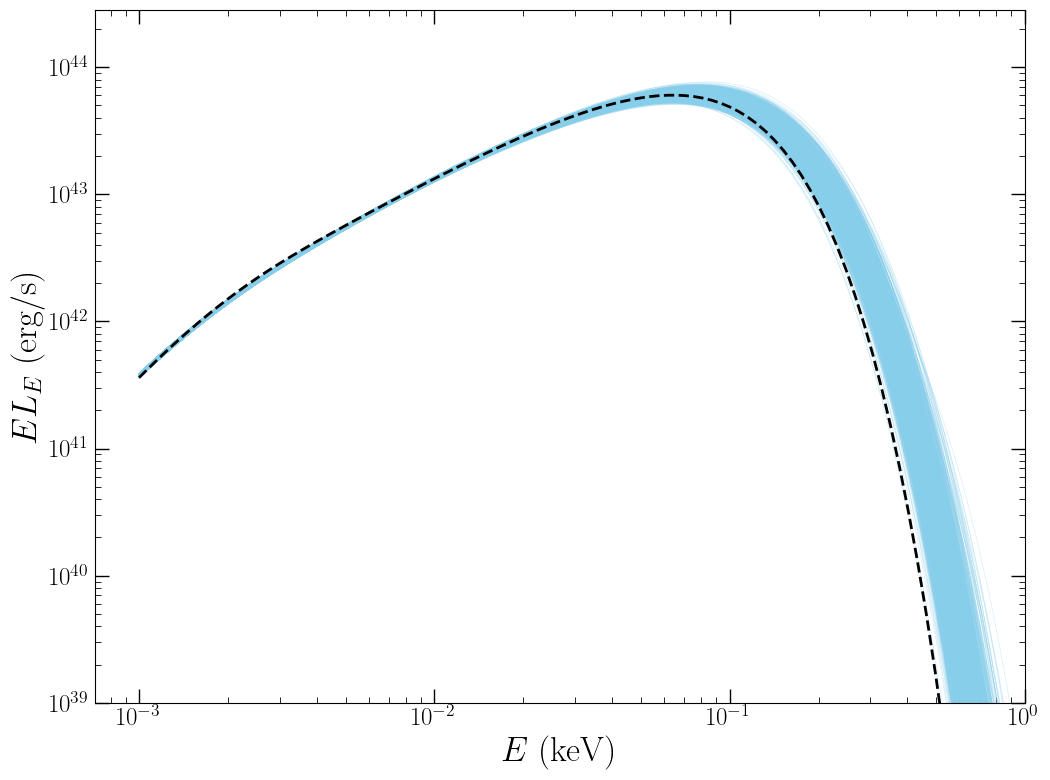}
    \caption{A set of 1600 observed disc spectra, generated from each of the disc temperature snapshots (see Fig. \ref{fig:example_snapshot}).  The black hole mass was set to $M = 2 \times 10^6 M_\odot$, and the average mass accretion rate $\dot M_0 = 6 \times 10^{20}$ kg/s, so as to produce typical $0.3-10$ keV X-ray luminosities typical of a tidal disruption event (a particularly relevant astrophysical system). It is clear that the fractional variability in the disc luminosity increases as a function of observing energy, and is particularly pronounced in the Wien tail. The conventional 1D disc theory spectrum for these parameters is shown by a black dashed curve. {Note that for the purposes of computing light curves the disc spectrum is evaluated above 1 keV, but this is not shown here. }   }
    \label{fig:example_spectra}
\end{figure}

In Fig. \ref{fig:example_spectra} we show an example series of observed spectra (plotted in terms of the spectral luminosity $E L_E$), as a function of observing energy $E$. Each curve corresponds to one of the 1600 disc snapshots recorded in the simulations of \cite{Turner23}. These spectral snapshots include the relativistic effects we discuss in section \ref{rel_sec}, and so represent our most sophisticated model.  The physical parameters of the system are chosen so that the system corresponds to a typical tidal disruption event system $M = 2 \times 10^6 M_\odot$, $\dot M_0 = 6 \times 10^{20}$ kg/s, with an average $0.3-10$ keV X-ray luminosity of $L \sim {\rm few} \times 10^{41}$ erg/s. Clearly, the stochastic nature of the disc temperature results in a stochastic disc spectrum, and this stochasticity is an inherent function of observing energy.  Namely, at low observing energies the two-dimensional stochastic structure of the disc temperature effectively averages out over the $2\pi$ azimuthal range of the disc, and the disc temperature is reasonably well described by the canonical one-dimensional thin disc theory (contrast the numerical spectra to the black-dashed classical theory prediction). 
However, at high observing frequencies (relative to the peak disc temperature), the spectrum is much more sensitive to the local physics of the inner accretion flow, and the fractional variability in the spectrum is much larger. This is especially true in the Wien tail of the spectrum, where order of magnitude spectral changes can be observed \citep[as predicted analytically;][]{MummeryBalbus22}. 

\begin{figure}
    \centering
    \includegraphics[width=\linewidth]{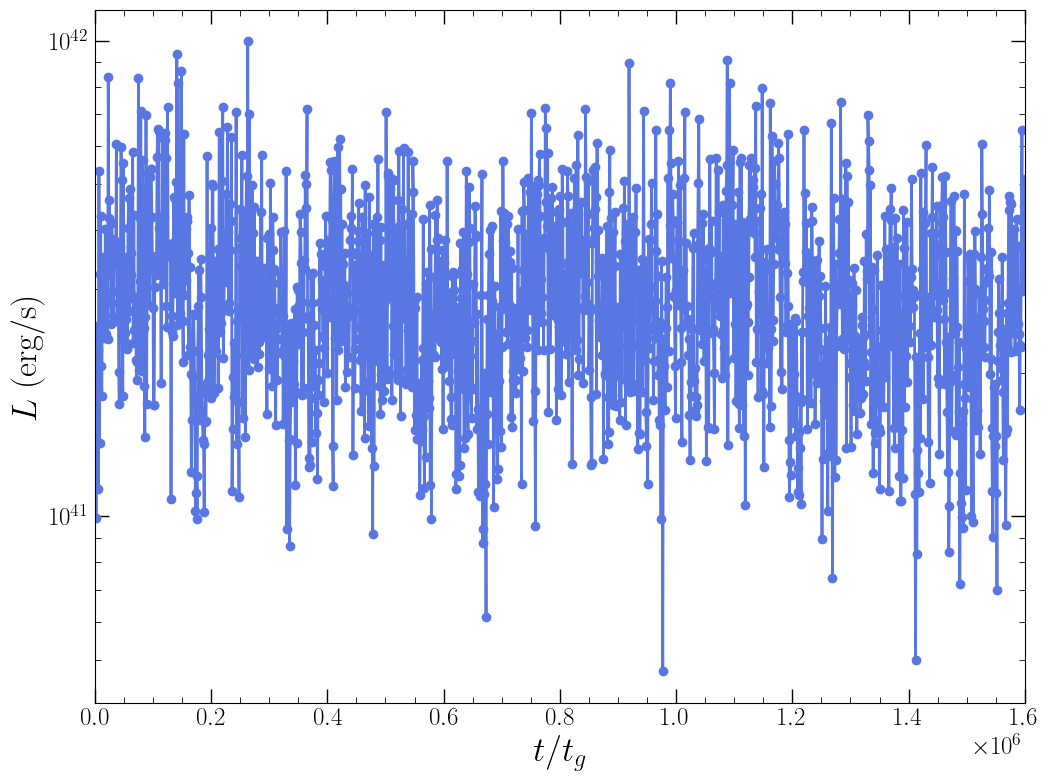}
    \includegraphics[width=\linewidth]{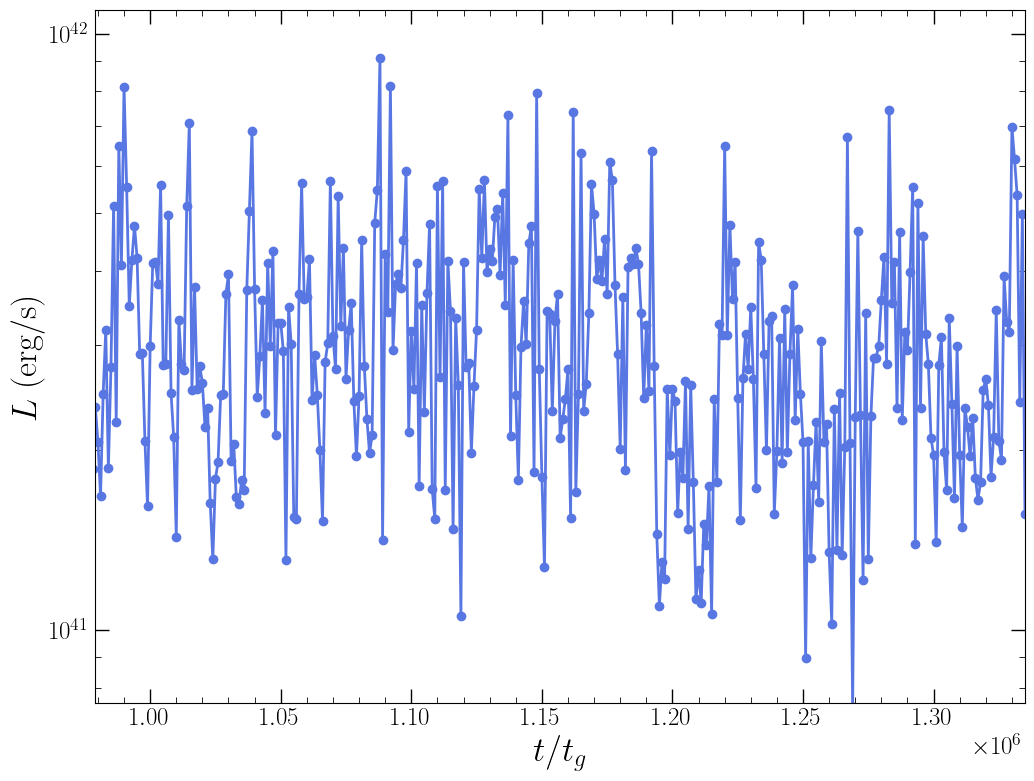}
    \caption{ Example lightcurves observed across $0.3-10$ keV, for the spectra depicted in Fig. \ref{fig:example_spectra}. Upper: the full lightcurve, while a temporal zoom-in is shown in the lower panel. For a black hole mass of $M = 2\times10^{6} M_\odot$, the gravitational timescale $t_g = GM/c^3 \simeq 10$ seconds, and so each point here is separated by $\Delta t = 10^3 t_g \sim$ 2 hours 45 minutes. Clearly, when observed in the Wien tail, order of magnitude X-ray luminosity fluctuations are possible on $\sim$ hour timescales.   }
    \label{fig:example_lightcurve}
\end{figure}

In addition, the luminosity observed across a band, defined by 
\begin{equation}\label{lum_int}
    L \equiv \int_{E_l}^{E_h} L_E \, {\rm d}E,
\end{equation}
can also be computed. In Fig. \ref{fig:example_lightcurve} we show the $0.3-10$ keV X-ray lightcurve computed by integrating the spectra in Fig. \ref{fig:example_spectra} across the X-ray bandpass (i.e., $E_l = 0.3$ keV, $E_h = 10$ keV). The lightcurve is notable for its large intrinsic variability, showing order of magnitude fluctuations in X-ray luminosity on timescales as short as the time between saved snapshots $\Delta t = 10^3 t_g$. Note that for typical tidal disruption event parameters the gravitational timescale is short $t_g = GM/c^3 \simeq 5 (M/10^6M_\odot)$ seconds. It is therefore likely that short ($\sim$ hour-day) timescale X-ray variability will be a feature of these events (we will discuss current observational evidence for such behaviour in later sections of this paper). 

It is clear that the variability of accretion flows at high photon energies has the potential to be leveraged as a probe of small length scale variability in accretion flows, particularly for tidal disruption event discs which fortuitously fall into this observational regime.   In the following section we introduce the analytical model of \cite{MummeryBalbus22}, which seeks to describe the probability density function of the high photon energy disc lightcurves. The purpose being to describe the variance seen in these systems, and link the observables to the local structure of the disc through a simpler (to analyse and compute) analytical model based on classical disc theory.  

\section{ Analytical analysis }\label{analy_sec}
\subsection{ The high-energy luminosity  probability density function }
As can be clearly seen in Fig. \ref{fig:example_spectra}, the fractional variability observed in the spectrum of an accretion disc is an increasing function of observing energy.   The large-variance properties of the Wien-tail was predicted analytically by \cite{MummeryBalbus22}, and can be intuitively understood as the exponential enhancement of disc temperature fluctuations (e.g., Fig. \ref{fig:example_snapshot}) from the $L \sim \exp(-E/kT)$ spectral shape of the Wien tail. 

This intuitive explanation can be placed on a firmer theoretical footing by treating the disc temperature $T(r, \phi, t)$ as a random variable, and exploring the implications of this variability for the spectrum (eq. \ref{newt_spec}) and disc luminosity (eq. \ref{lum_int}). Both of these variables now must of course be considered random variables (a randomness they inherit from the underlying turbulent disc properties). The analysis proceeds by deriving analytical approximations for the two spectral integrals (eqs. \ref{newt_spec}, \ref{lum_int}), before folding through the randomness of the temperature field. 

In \cite{MumBalb21a} it was shown that equations (\ref{newt_spec}) and (\ref{lum_int}) can be evaluated by performing a Laplace expansion about the hottest region in the disc.  The physical insight here is that only the very hottest disc regions contribute to the spectrum (or luminosity) in the Wien-tail, as the flux from all other disc regions is exponentially suppressed \citep[see][for further discussion]{Balbus14}. 
 
Performing the Laplace expansion and then integrating results in the following expression for the disc luminosity (eq. \ref{lum_int})
\begin{equation}\label{MB}
L  = \frac{16 \pi^2  \chi_1 E_l^4}{c^2 h^ 3} R_p^2 \cos i \left(\frac{k {T}_p}{E_l} \right)^\eta \exp\left(- \frac{E_l}{k {T}_p} \right) .
\end{equation} 
Here we have defined $T_p$ the hottest temperature in the accretion disc. The radius $R_p$ corresponds to the radial location of the peak temperature, while the constant $\eta$ depends on the inclination angle of the disc and the disc's inner boundary condition, and is limited to the range $3/2 \leq \eta \leq  5/2$. The constant $\chi_1 \simeq 2.19$ for classical thin disc theory, although this assumes an axi-symmetric one-dimensional disc temperature profile.  As this is not the case in our numerical simulations, the parameter $R_p$ may differ by an order unity factor from the true radial scale of the disc temperature. 

This result clearly highlights an important effect of observing systems in the Wien-tail: only a small region of the disc contributes to leading order (the hottest disc temperature is the only disc variable which appears). This means that local turbulence will play an oversized role in Wien-tail luminosity variance, as the observable is sensitive to the small scale structure of the disc. While this may make such observations difficult to fit using traditional approaches, it does open up the possibility of probing small scale disc turbulence with observations. 

It should be noted that equation (\ref{MB}) was derived in \cite{MumBalb21a} assuming a temperature profile which depend on only one (radial) spatial dimensional (e.g., $T(r, t)$). Such an approximation is not required, and identical temperature scaling relationships (i.e., $L \propto T^\eta \exp(-E/kT)$) can be derived for two dimensional temperature profiles \citep[as shown in][]{Balbus14, MumBalb20a}, with differing (but largely physically irrelevant) prefactors in this two-dimensional case.

The mathematical question is then, if the high energy luminosity is described by a (dimensionless) function of the form 
\begin{equation}\label{dimX}
Y = X^\eta \, \exp(-1/X) ,
\end{equation}
and $X$ is a random variable, how is $Y$ distributed? 
In the above expression $Y = L/L_0$ is a normalised luminosity,  $X = k T_p/E_l$ a normalised temperature, and  $L_0$ is a  disc temperature independent constant defined by
\begin{equation}\label{constdef}
L_0 \equiv {16\pi^2 \chi_1 \over c^2 h^3 } E_l^4 R_p^2 \cos i.  
\end{equation}
The parameter $E_l$ is the lower bandpass energy from equation (\ref{lum_int}).  The answer to this question is \citep{MummeryBalbus22}
\begin{equation}\label{p_gen}
    p_Y(y) =  { \eta^{-2} \, \, y^{-1} \over W(z)\left(1 + W(z)\right) } \,\, p_X\left( \left[ \eta W\left(z \right) \right]^{-1}\right),
\end{equation}
where $p_X$ is the probability density function from which the random variable X (in this case the disc temperature) is drawn 
\begin{equation}
    X \sim p_X(x), 
\end{equation}
and $W$ is the Lambert W function  \citep{Corless96},  which is defined as the solution $w$ of the following equation 
\begin{equation}\label{lambdef}
w e^w = z \rightarrow w \equiv W(z),
\end{equation}
and $z \equiv \eta^{-1} y^{-1/\eta}$. 

\cite{MummeryBalbus22} made the additional assumption that the probability density function of temperature fluctuations followed a log-normal profile \citep[a result with much observational support, e.g.,][]{Uttley05}. The log-normal distribution has the following form 
\begin{equation}\label{pln}
p_X (x; \mu_N, \sigma_N) = {1\over \sqrt{2\pi}\sigma_N x }\exp (-[\ln(x)-\mu_N]^2/2\sigma_N^2),
\end{equation}
where this function is valid only for $x > 0$ (positive temperatures), and $\mu_N$ and $\sigma_N^2$ are the mean and variance of the underlying normal distribution (these will be related to the mean $\mu_T$ and variance $\sigma_T^2$ of the temperature distribution shortly). Under this assumption the X-ray luminosity should be distributed according to 
\begin{multline}\label{xray_dist}
p_Y(y; \eta, \mu_N, \sigma_N) = {\eta^{-1} y^{-1} \over \sqrt{2\pi} \sigma_N \left(1 + W(z) \right) } \times \\
\exp\left[- \left(W(z) + \eta^{-1}\ln(y) - \mu_N\right)^2 \Big/ 2\sigma_N^2\right] . 
 \end{multline}
This model may now be tested with the viscous hydrodynamic disc models of \cite{Turner23} {or with observations of astrophysical systems. If temperature fluctuations in astrophysical systems are not log-normally distributed then eq. (\ref{xray_dist}) will not accurately describe the luminosity variability, but the more general result (equation \ref{p_gen}) will still hold. The modelling of different temperature distributions in equation (\ref{p_gen}) could be leveraged as a future test of one of the key predictions of the theory of propagating fluctuations (the log-normality of emission).  }

Recall that the true luminosity variable $L$ is related to the variable $Y$ by a simple scaling $L = L_0 Y$, and that $L_0$ depends on the system parameters only through the black hole mass $M$ (which sets the radial scale $R_p$ in equation [\ref{constdef}]). 
The high energy luminosity distribution is therefore a four-parameter distribution which may be found by combining equation (\ref{xray_dist}) with the definition of $L_0$ (equation [\ref{constdef}]): 
\begin{equation}    
p_{L}(l;R_p, \eta, \mu_T, \sigma_T) = L_0^{-1} \, p_Y(y = l/L_0; \eta, \mu_N, \sigma_N) ,
\end{equation}
where the mean ($\mu_N$) and variance ($\sigma_N^2$) of the underlying normal distribution which appear in equation (\ref{xray_dist}), are related to the  mean ($\mu_T$) and variance ($\sigma_T^2$) of the temperature distribution by 
\begin{align}
\mu_N &= \ln\left({k \mu_T \over E_l} {\mu_T \over \sqrt{\mu_T^2 + \sigma_T^2}}\right) ,\label{normmu} \\
\sigma_N^2 &= \ln\left(1 + {\sigma_T^2 \over \mu_T^2}\right) . \label{varmu}
\end{align}

\subsection{ Modelling the disc variability }
\begin{figure}
    \centering
    \includegraphics[width=\linewidth]{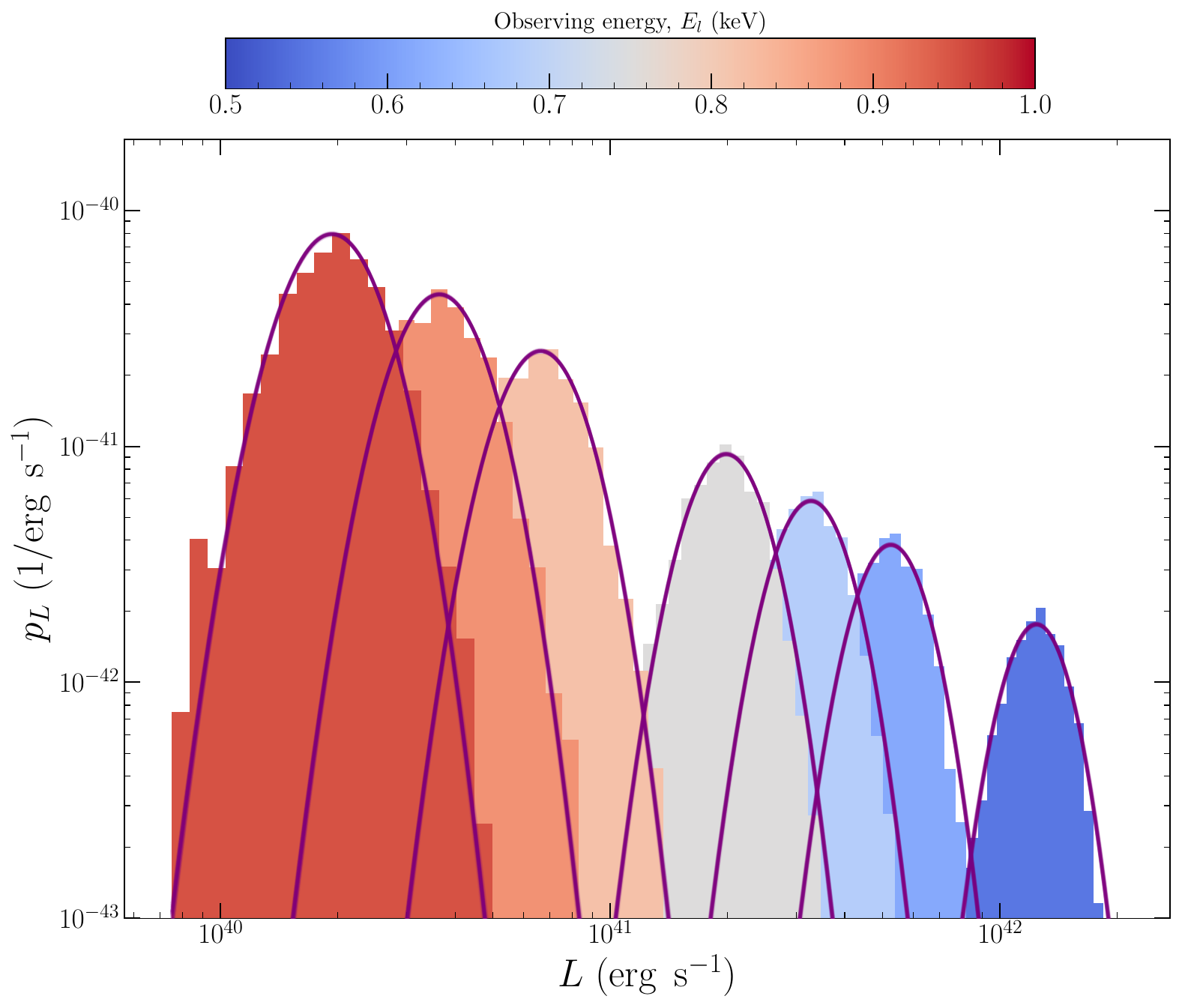}
    \caption{The numerical (coloured histograms) and theoretical (purple curves) distributions of the X-ray lightcurves. Each lightcurve was generated from the same disc (with $M = 10^6 M_\odot, \dot M_0 = 6 \times 10^{21}$ kg/s), but was observed with a different lower bandpass energy $E_l$. The blue (brightest) histogram represents the lowest observing energy, while the red (dimmest) shows the highest. These different bandpass energies were chosen so that the average luminosity across the bands varied by a factor $\sim 100$. The purple curves are simultaneous fits of the analytical distribution (eq. \ref{xray_dist}) to the numerical data; the fit is excellent.    }
    \label{fig:fit_to_lc}
\end{figure}

\begin{figure}
    \centering
    \includegraphics[width=\linewidth]{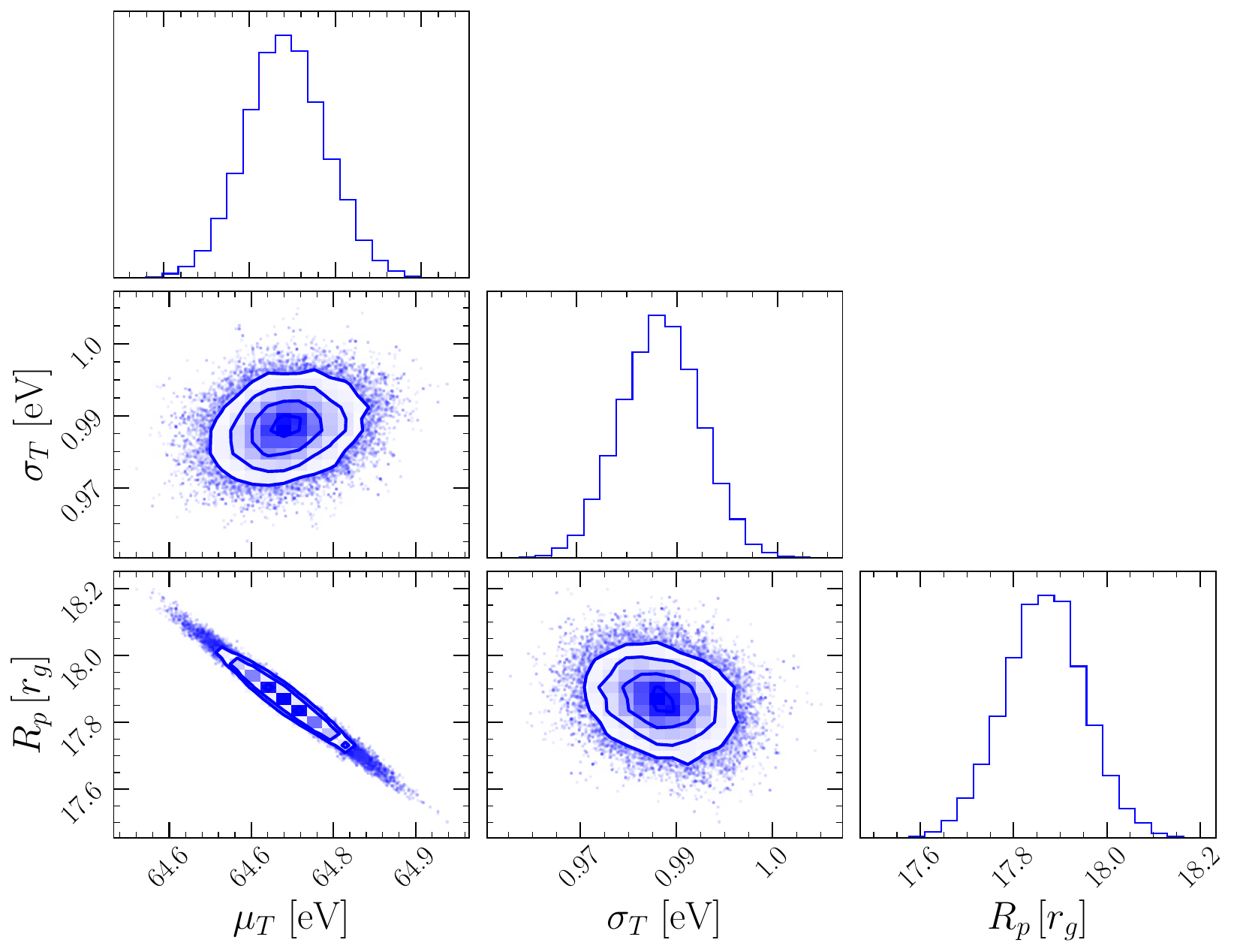}
    \caption{The correlations between the parameters of the X-ray luminosity probability density function (eq. \ref{xray_dist}), and their posterior distributions, for the fit to the numerical distributions shown in Fig. \ref{fig:fit_to_lc}. The fractional temperature variance $\sigma_T/\mu_T$ is of order $\sim 1.5\%$, despite the X-ray lightcurves showing significant (factor $\sim$ few) variability.    }
    \label{fig:corner_to_lc}
\end{figure}

The question we pose in this section is whether the emergent high-energy luminosity of the stochastic-viscosity hydrodynamic disc model of \cite{Turner23} (specifically the model with ${l_c=H=0.1r}$) can be described by the analytical results of \cite{MummeryBalbus22}. 

The luminosity probability density function (eq. \ref{xray_dist}) is a function of four free parameters $\Theta = (\eta, \mu_N, \sigma_N, L_0)$. For a given X-ray lightcurve (e.g., Fig. \ref{fig:example_lightcurve}), the likelihood of observing that lightcurve, given the set of free parameters of the theory, is given by definition by the expression 
\begin{equation}\label{likelihood}
    {\cal L}(\Theta) = \prod_{j = 1}^{N} p_L(l_j; \Theta), 
\end{equation}
where $\left\{l_j\right\}$ is the list of $N$ luminosity values. This explicitly assumes that each observed luminosity is independent of all previous luminosities; an assumption which is only true if the luminosities are observed at times separated by a period longer than the typical driving timescale of the turbulence. For the models considered here, the driving timescale in the (relevant) inner regions is of order $\sim 150 t_g$ (see eq. \ref{timescale}), whereas we ``observe'' on timescales of $1000t_g$, and this approximation is therefore valid. 

To test the ability of the analytical result (eq. \ref{xray_dist}) to reproduce the numerical light curves produced by the model of \cite{Turner23}, we generate X-ray lightcurves $L(t)$ for a disc model with $M = 10^6 M_\odot$ and $\dot M_0 = 6 \times 10^{21}$ kg/s, corresponding to typical tidal disruption event values (with an average $0.3-10$ keV X-ray luminosity of order $L_X \sim 10^{42}$ erg/s). {Note that the actual values of the parameters here are unimportant, owing to the dimensionless unit system that {\tt PLUTO} employs. As the two dimensional temperature structure is the same for each run (with a given set of turbulence parameters) shifting the temperature by 
\begin{equation}
    T(r, \phi, t) \to T(r, \phi, t) \left[\dot M_0 / M^2\right]^{1/4}    
\end{equation} 
will result in identical variability structure (about a different mean amplitude) when observed at an energy}
\begin{equation}
    E_l \to E_l \left[\dot M_0 / M^2\right]^{1/4}.
\end{equation} 
We  generate lightcurves for a range of values of $E_l$ (but fixed disc parameters), chosen so that the typical average luminosities of the different bands span a factor of 100. The smallest value of $E_l$ is 0.5 keV, and the larger $E_l$ values are all then equally spaced up to $E_l = 1$ keV. 

To model these lightcurves, we fit the temperature parameters $\mu_T$ and $\sigma_T$, and the disc size parameter $R_p$ by maximising the log-likelihood 
\begin{equation}
    \log {\cal L}(\Theta) = \sum_{{\rm energies}, i} \, \sum_{j = 1}^{N} \log p_L(l_j; \Theta; E_{l, i}), 
\end{equation}
using Monte-Carlo Markov chain techniques \cite{EMCEE}. The results are displayed in Figs. \ref{fig:fit_to_lc} \& \ref{fig:corner_to_lc}. In Fig. \ref{fig:fit_to_lc} we display the numerical (coloured histograms) distribution of disc luminosities (e.g., histograms of the disc lightcurves like those displayed in Fig. \ref{fig:example_lightcurve}). As purple solid curves we display a fit to the analytical probability density function (eq. \ref{xray_dist}), with one set of disc parameters $(\mu_T, \sigma_T, R_p)$ fit to all of the distributions simultaneously. We fix $\eta = 5/2$, which was found to best describe the numerical data. Physically this value of $\eta$ corresponds to a temperature profile with sharp gradients, which is perhaps unsurprising given the temperature structure seen visually (Fig. \ref{fig:example_snapshot}), see \cite{MumBalb21a} for further discussion on the physics of $\eta$.  
It is clear that the analytical distribution (eq. \ref{xray_dist}) well describes the numerical results. 

This is an important result. It is not obvious that the variable luminosity resulting from a two-dimensional viscous-hydrodynamics simulation, with a stochastic viscosity, would be well described by simple analytic disc theory. {As can be seen in eq. \ref{eq:diss} the disc temperature is a non-linear function of viscosity, density and the velocity gradients of the flow, all of which are implicitly dependent on the stochastic variable $\beta$. The fact that this complex behaviour can be quite accurately described by thin disc theory coupled to a log-normal temperature profile is non-trivial, particularly as the local accretion rate and local temperature are no longer tightly correlated in these simulations \citep[see Figure 20 of][for more details]{Turner23}. }  The fact that this {simple description works well} opens up the possibility of using the theory (eq. \ref{xray_dist}) to probe the properties of accretion disc turbulence, or alternatively help explain some of the variability of those sources observed in the relevant limit of $E_l \gg kT_p$. 

The values of (and correlations between) the best fitting parameters of the luminosity distribution are displayed in Fig. \ref{fig:corner_to_lc}. While fitting we normalise the radial coordinate by $r_g = GM/c^2$, where $M$ is the (exact) black hole mass variable used in generating the lightcurves, so that $R_p$ represents a typical (dimensionless) length scale of the peak disc temperature. 
An extremely important result is that the fractional variability of the disc temperature is only of the percent level $(\sigma_T/\mu_T \simeq 1.5\%)$, despite the X-ray variability of the light curve being at the factor $\sim 2$ level for lower observing frequencies (blue histogram Fig. \ref{fig:fit_to_lc}), and the factor $\sim 5$ level for higher observing frequencies (red histogram, Fig. \ref{fig:fit_to_lc}). This highlights the exponential amplification of disc temperature variability when observing in the Wien-tail. 

\begin{figure}
    \centering
    \includegraphics[width=\linewidth]{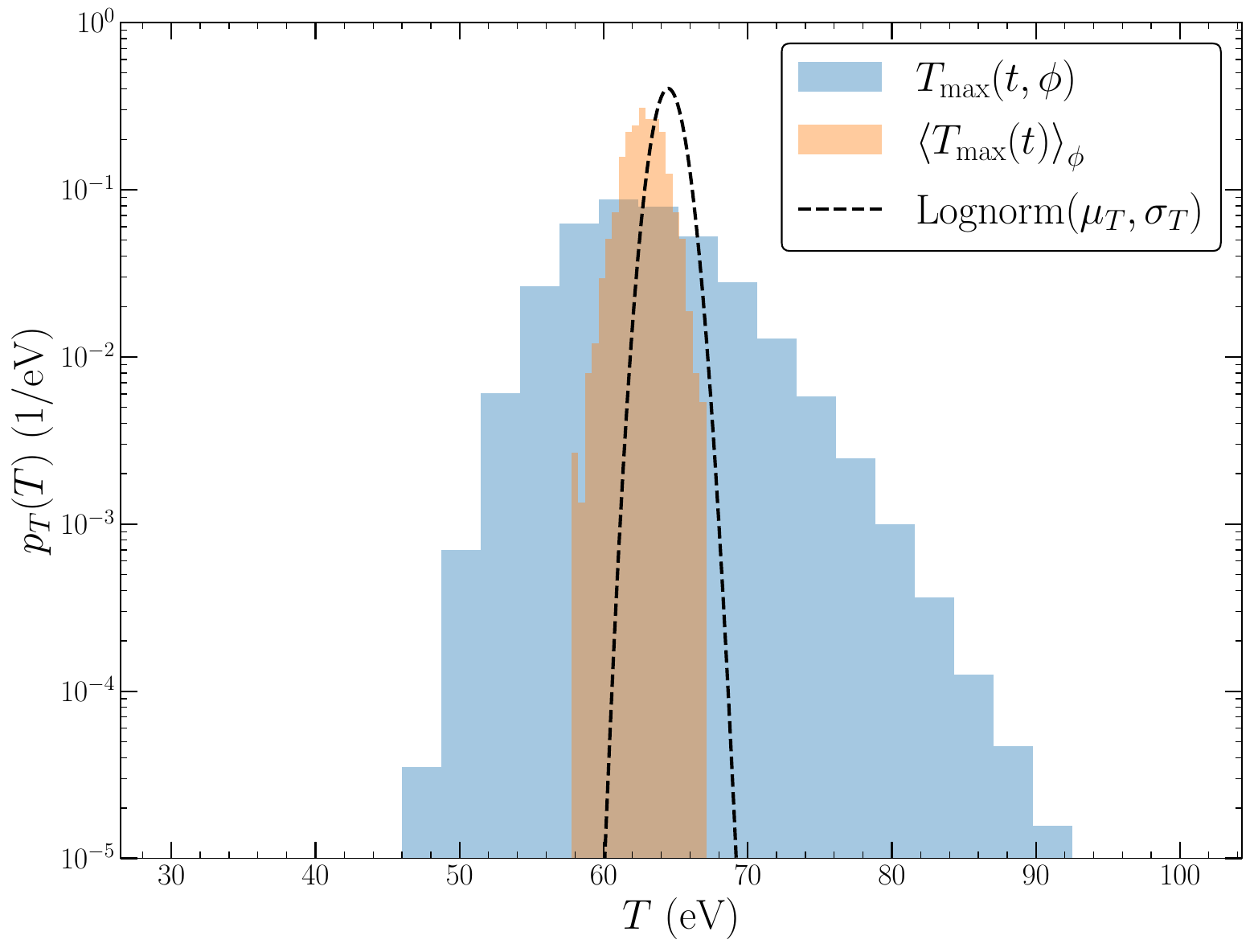}
    \caption{ The distributions of the fixed azimuth maximum temperatures $T_{\rm max}$, and the angle averaged maximum temperature $\left\langle T_{\rm max}(t) \right\rangle_\phi $ (produced with $M = 10^6 M_\odot, \dot M_0 = 6 \times 10^{21}$ kg/s, see text). The small scale temperature variability (i.e., the one probed by $T_{\rm max}(\phi, t)$) is significantly more variable than the angle-averaged temperature scale $\left\langle T_{\rm max}(t) \right\rangle_\phi$.  However, as can be seen by comparing the fitted distribution (black dashed curves) to the physical distributions, the light curve variability is (at least in a Newtonian system) dominated by the variability of the angle-averaged temperature profile. }
    \label{fig:temp-dists}
\end{figure}

It is interesting to compare the fitted distribution of the peak disc temperature (i.e., a log-normal distribution with mean $\mu_T$ and standard deviation $\sigma_T$) with the physical properties of the disc system. Clearly (e.g., Fig. \ref{fig:example_snapshot}) the disc temperature is a complicated two-dimensional random field. We can define two different ``peak disc temperatures'', the first being the maximum temperature in the disc at a given azimuth at a fixed time (the rationale here being that the angular structure probes the turbulent variation, as the radial temperature dependence is driven principally by the gravitational energy liberation)
 \begin{equation}
     T_{\rm max}(\phi, t) = \max_r \left\{ T(r, \phi, t) \right\} ,
 \end{equation}
while we can also define an angle-averaged maximum temperature 
\begin{equation}
    \left\langle T_{\rm max}(t) \right\rangle_\phi \equiv {1 \over 2\pi} \int_0^{2\pi} T_{\rm max}(\phi, t) \, {\rm d}\phi ,
\end{equation}
which is likely more closely related to the classical disc theory prediction. We plot the two distributions in Fig. \ref{fig:temp-dists}, which highlights the difference in these two values and the structure of the accretion flow. The small scale temperature variability (i.e., the one probed by $T_{\rm max}(\phi, t)$) is significantly more variable than the angle-averaged temperature scale $\left\langle T_{\rm max}(t) \right\rangle_\phi$. Indeed, the small scale temperature structure varies by a factor of $\sim 2$ over the entire simulation, while the averaged temperature shows significantly smaller variability. 

Also shown in Fig. \ref{fig:temp-dists} is the best fitting temperature distribution fit to the X-ray lightcurves observed from this system (Fig. \ref{fig:fit_to_lc}), displayed by black dashed lines. Clearly, the light curve variability is (at least in a Newtonian system) dominated by the variability of the angle-averaged temperature profile. However, the properties of observing in the Wien tail mean that the inferred temperature profile is shifted slightly to higher temperatures, as hotter disc regions contribute exponentially more to the light curves. 

It is important to note that while we have generated different luminosity distributions (in Fig. \ref{fig:fit_to_lc}) by varying the observing energy $E_l$, there is of course an exact symmetry here with varying the mean disc temperature $\mu_T$ (i.e., by changing the mass accretion rate, or black hole mass) at fixed observing energy. What is important \citep[as stressed in][]{MummeryBalbus22} is the ratio between the average peak disc temperature and the observing energy $k\mu_T/E_l$. A $\sim 1\%$ temperature variability can, if observed deep enough into the Wien-tail, result in order of magnitude luminosity variability. 

\begin{figure}
    \centering
    \includegraphics[width=.95\linewidth]{andyfigs/example_snapshot.png}
    \includegraphics[width=.95\linewidth]{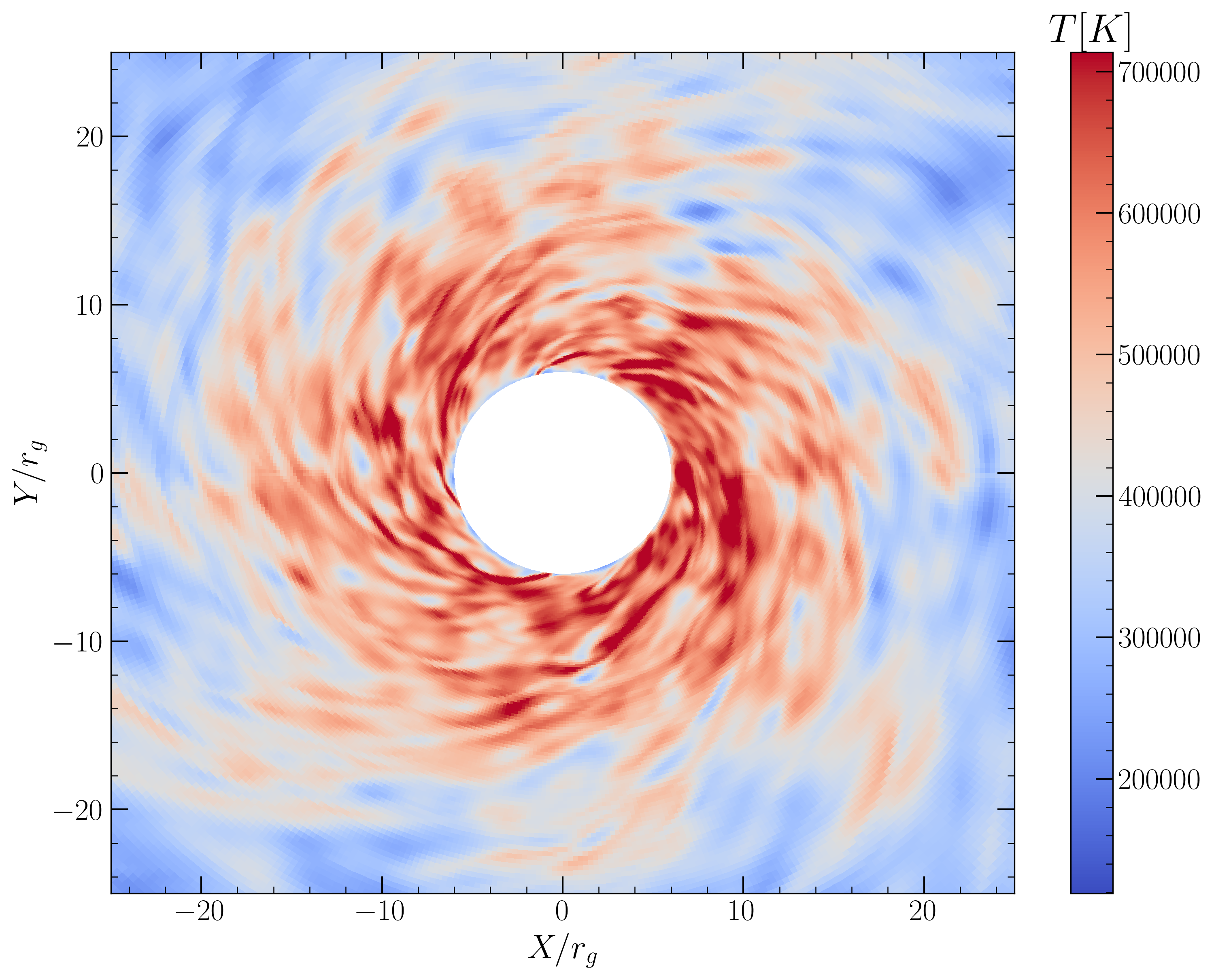}
    \includegraphics[width=.95\linewidth]{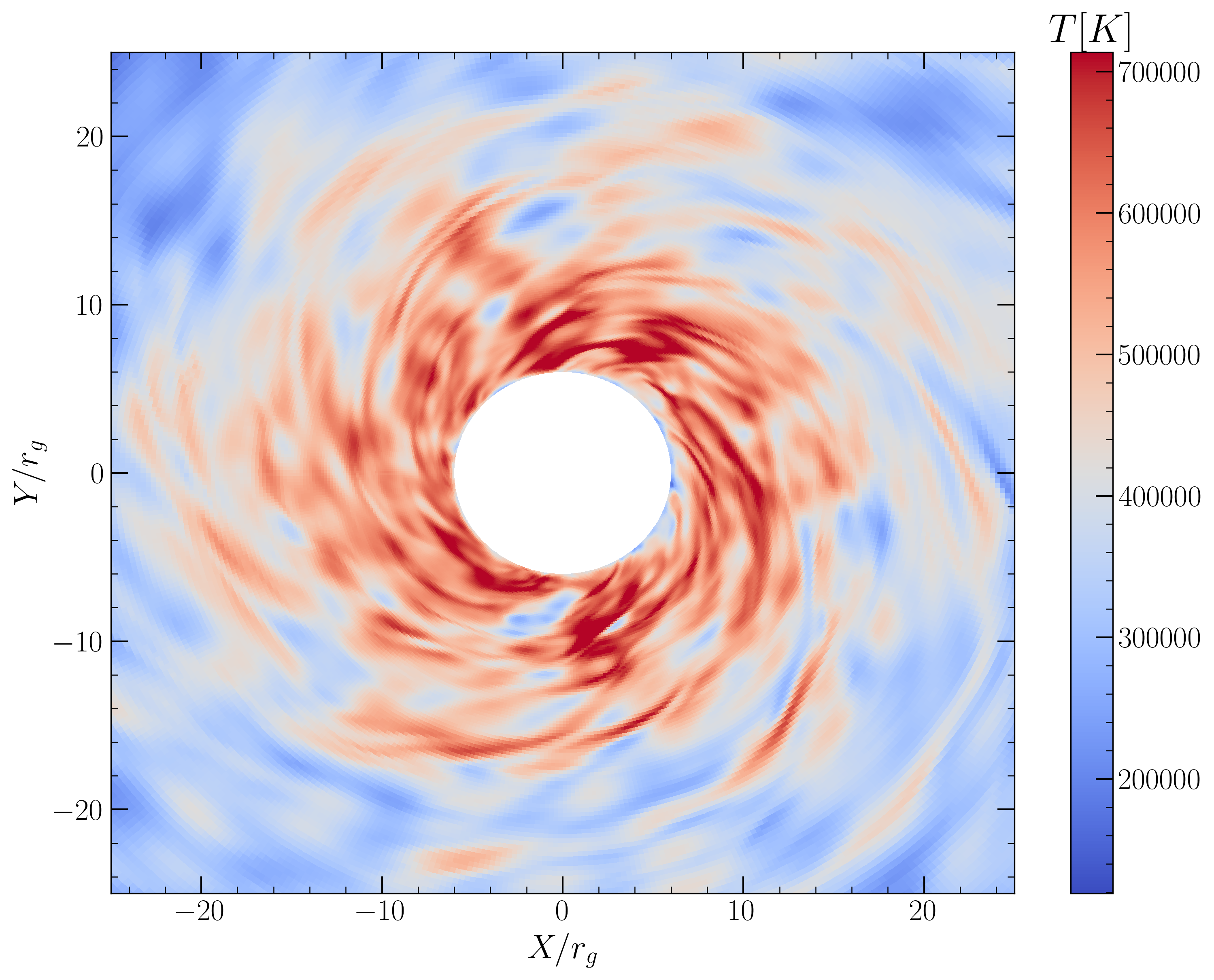}
    \caption{The effects of increasing the coherence length of the stochastic viscosity prescription on the disc temperature profile: {upper panel $l_c/r = 0.1$, middle panel $l_c/r = 0.15$, lower panel  $l_c/r = 0.2$}. As the coherence length of the viscosity is increased (from upper to lower panels) the temperature variability becomes increasingly structured on large scales, and clear spiral structures are visible. This results in larger amplitude luminosity variability, particularly when observed in the Wien-tail.  }
    \label{fig:disc_phys}
\end{figure}

\subsection{ The effects of changing the disc turbulence physics } \label{sec:turb_physics}
As the analytical model of \cite{MummeryBalbus22} reproduces well the full numerical stochastic-viscosity hydrodynamics models of \cite{Turner23}, we now use this theory to examine the effects different assumptions about the properties of the disc turbulence have on the observed lightcurves of accretion flows at high photon energies. 

There are three principal components of the viscous parameterisation which can be modified: the driving timescale $1/\omega$; the amplitude of variability $\sqrt{\left\langle \beta^2\right\rangle}$; and the coherence length of the variability $l_c$.   Two of these, the timescale and amplitude of variability, result in relatively trivial (though important) modifications to the results. Namely, modifying the amplitude of variability $\sqrt{\left\langle \beta^2\right\rangle}$ simply scales the overall temperature (and therefore luminosity) variance. This is particularly important in the Wien-tail, owing to the exponential amplification of temperature fluctuations. Similarly, modifying the driving timescale $1/\omega$ simply scales the timescale of the peaks and troughs of the lightcurve flaring (e.g., Fig. \ref{fig:example_lightcurve}), without substantially modifying the amplitude of variability. As the canonical values chosen for these two parameters are selected with reference to MHD simulations of discs \citep[][see further discusion in \citealt{Turner23}]{Hogg16}, we do not modify them further in this analysis. 

The effects of modifying the coherence length of the stochastic viscosity parameterisation $l_c$ is less obvious, and physically interesting to examine. In real discs, the spatial coherence length scale $l_c$ is likely set (to first order) by the height $H$ of the disc. The idea being that turbulent eddies will have a size given approximately by the disc thickness, and certainly will not substantially exceed this value. \cite{Turner23} presented simulations which varied $H$ (but kept $l_c=H$) and also ones which kept ${\cal H}=0.1$ but varied $l_c$ independently. For the full details on these simulations and how the coherence length is modified see \citet{Turner23} (particularly Appendix A)\footnote{Note that the notation used in \citet{Turner23} is slightly different from that used here. There, the coherence length is defined relative to the aspect ratio of the disc through a coherence factor as ${l_c/r=f_{\rm coh}{\cal H}}$.}. They found that increasing the coherence length (whether or not this was accompanied by a change in $H$) increased the overall variability of their disc simulations. Therefore thicker discs (which are those which we would expect to be more coherent) will likely be more variable.

In order to see how this result feeds through into the high-photon-energy variability, we now consider three simulations from \citet{Turner23}. Along with the fiducial simulation we have used up until now (with $l_c/r={\cal H}=0.1$) we consider two more which keep ${\cal H}=0.1$ but have $l_c/r=0.15$ and $0.2$ respectively. While we are motivated here by considering the effect of disc thickness on variability, we have here chosen simulations with the same ${\cal H}$. In the vertically integrated simulations of \citet{Turner23}, changing ${\cal H}$ changes the inflow velocity for a given accretion rate (since a larger value of ${\cal H}$ increases the viscosity and thus lowers the surface density and increases the inflow velocity; see eq. \ref{steady_state}). The choice of keeping $\cal H$ constant is therefore made to allow for easier interpretation of our results, as it ensures that the only difference between the simulations is the value of $l_c$.  Note however that, as mentioned earlier, we would expect very similar results if we considered the simulations where $\cal H$ was changed along with $l_c/r$, though this is typically harder to simulate.

In Fig. \ref{fig:disc_phys}, we show temperature snapshots for the three simulations, ordered top to bottom by increasing coherence length. Each disc temperature snapshot was made with the typical TDE parameters of $M = 2\times10^6 M_\odot$, $\dot M_0 = 6 \times 10^{20}$ kg/s. It is relatively clear to see the effects of increasing the coherence length on the resulting disc temperature profile: unsurprisingly the disc temperature is coherent over much larger scales for longer $l_c$ (i.e. the increased coherence length in the viscosity is translated directly into that for the temperature), and the fluctuations resemble more macroscopic sheared-out spiral structures, rather than very short length scale fluctuations.

\begin{figure}
    \centering
    \includegraphics[width=\linewidth]{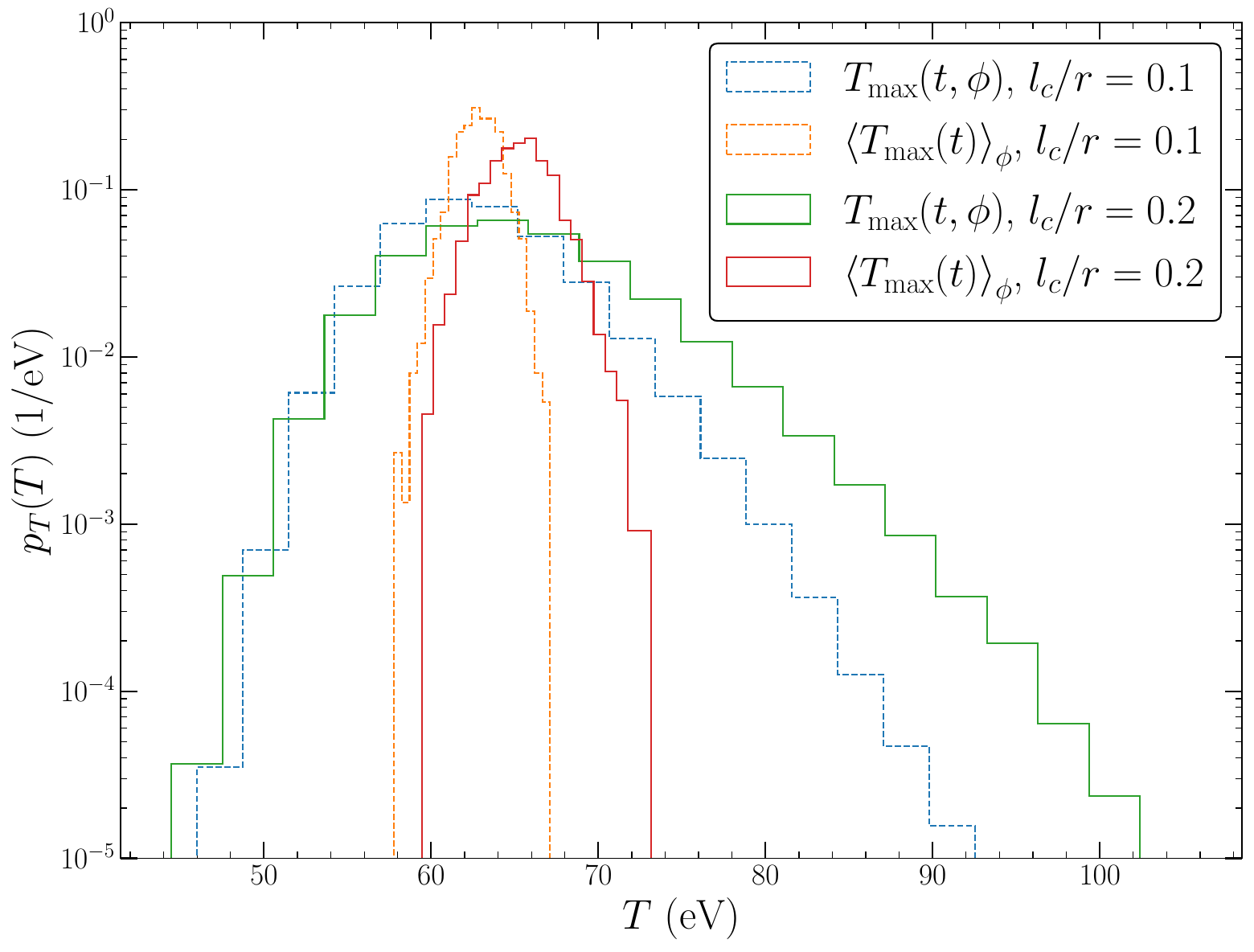}
    \caption{A comparison between the fixed azimuth and angle-averaged peak temperature profiles of discs with different coherence lengths (displayed in legend; dashed curves have $l_c/r = 0.1$, solid curves have $l_c/r=0.2$; both produced with $M = 10^6 M_\odot, \dot M_0 = 6 \times 10^{21}$ kg/s). Increasing the coherence length both shifts the peak temperature distribution to higher temperatures, but also increases the variance in this distribution. This change becomes clearly observable in the light curves (e.g., Fig. \ref{fig:lc_double}).   }
    \label{fig:temp_dist_double_length}
\end{figure}

Quantitatively, increasing the coherence length both shifts the peak temperature distribution to higher temperatures, but also increases the variance in this distribution.  This can be seen clearly in Fig. \ref{fig:temp_dist_double_length}, where we display the fixed azimuth maximum temperatures $T_{\rm max}$, and the angle averaged maximum temperature $\left\langle T_{\rm max}(t) \right\rangle_\phi $ distributions, for both $l_c/r = 0.1$ and $l_c/r = 0.2$. 

The physical reason for this change is spelled out in detail in \cite{Turner23}; a disc with larger coherence length has fewer ``independent'' disc regions in the inner edge, with of order $N \sim 2\pi r_I / l_c$ independent disc regions near to the ISCO. As small scale disc structure is more highly variable (e.g., Figs. \ref{fig:example_snapshot}, \ref{fig:temp-dists}), the smaller the number of independent regions averaged over, the larger the variance in the temperature distribution. 

\begin{figure}
    \centering
    \includegraphics[width=\linewidth]{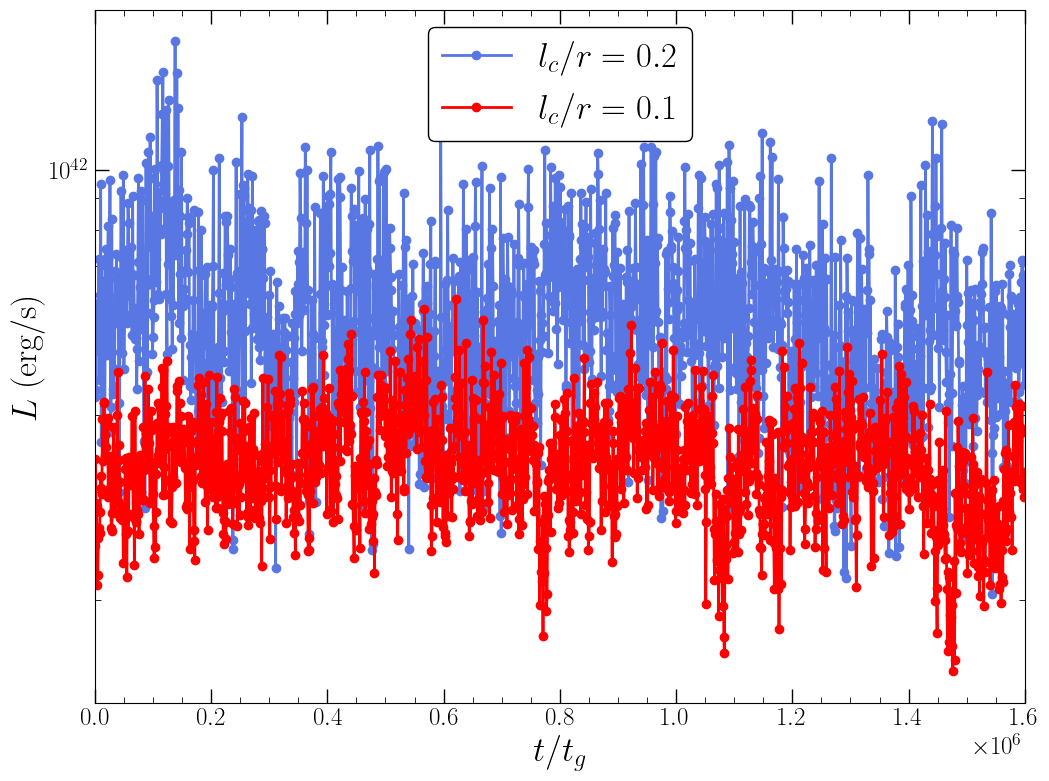}
    \includegraphics[width=\linewidth]{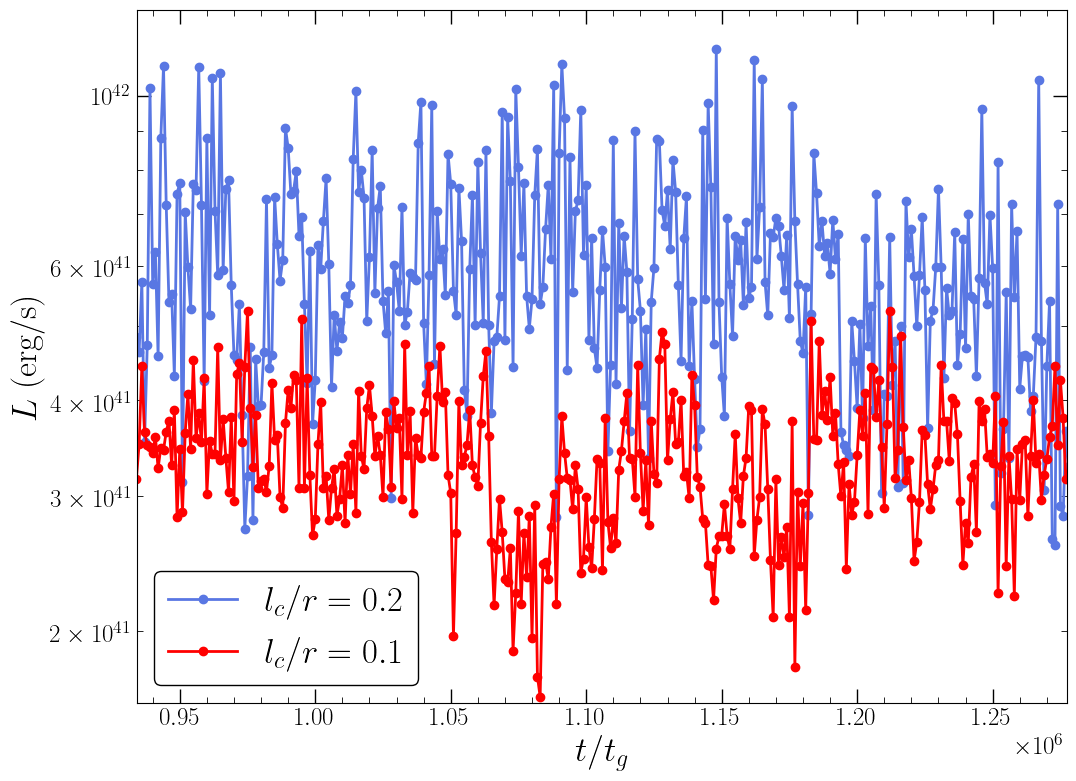}
    \caption{Example lightcurves for discs of different variability coherence lengths $l_c/r$ (displayed on plot). Upper: the full lightcurve, while a temporal zoom-in is shown in the lower panel. The discs and black holes of the two simulations were otherwise identical (with the same parameters as Fig. \ref{fig:fit_to_lc}).  Both the mean and variance of the disc luminosity are strongly effected by changing the coherence length of the disc variability, with both increasing with $l_c$.  Identifying $l_c$ with the height of the disc suggests that thicker discs will be both brighter and more variable.}
    \label{fig:lc_double}
\end{figure}

This modified variability coherence scale  has a pronounced effect on the Wien-tail lightcurves observed from the system. In Fig. \ref{fig:lc_double} we show the lightcurves for two discs of different variability coherence lengths $l_c/r = 0.1$ (red) and $0.2$ (blue). The discs and black holes of the two simulations were otherwise identical (with the same parameters as Fig. \ref{fig:fit_to_lc}).  Both the mean and variance of the disc luminosity are modified by changing the coherence length of the disc variability, with both increasing with $l_c$. If we identify $l_c$ with the height of the disc, this result suggests that thicker discs will be both brighter and more variable. This may have important observational implications.

\begin{figure}
    \centering
    \includegraphics[width=\linewidth]{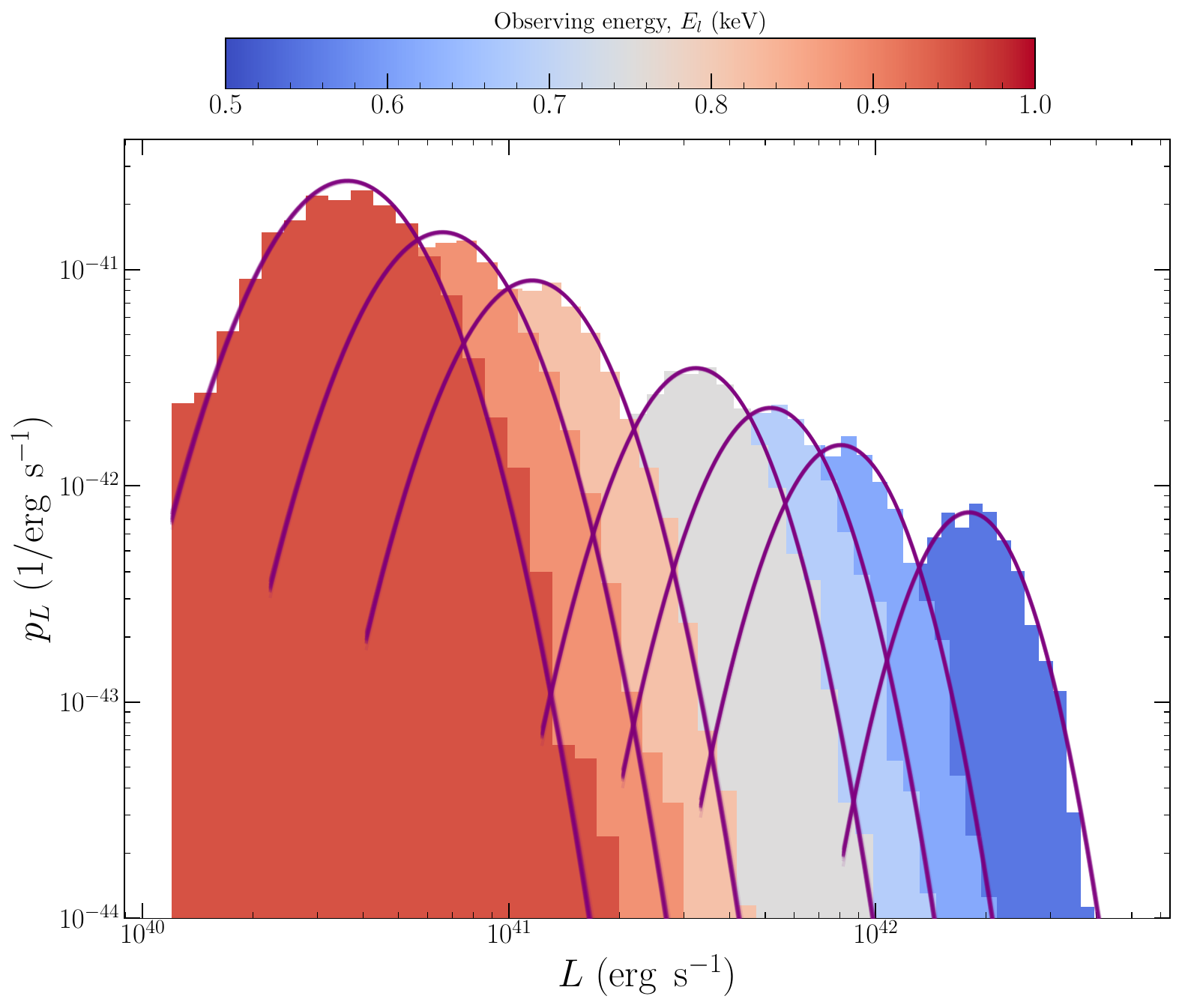}
    \includegraphics[width=\linewidth]{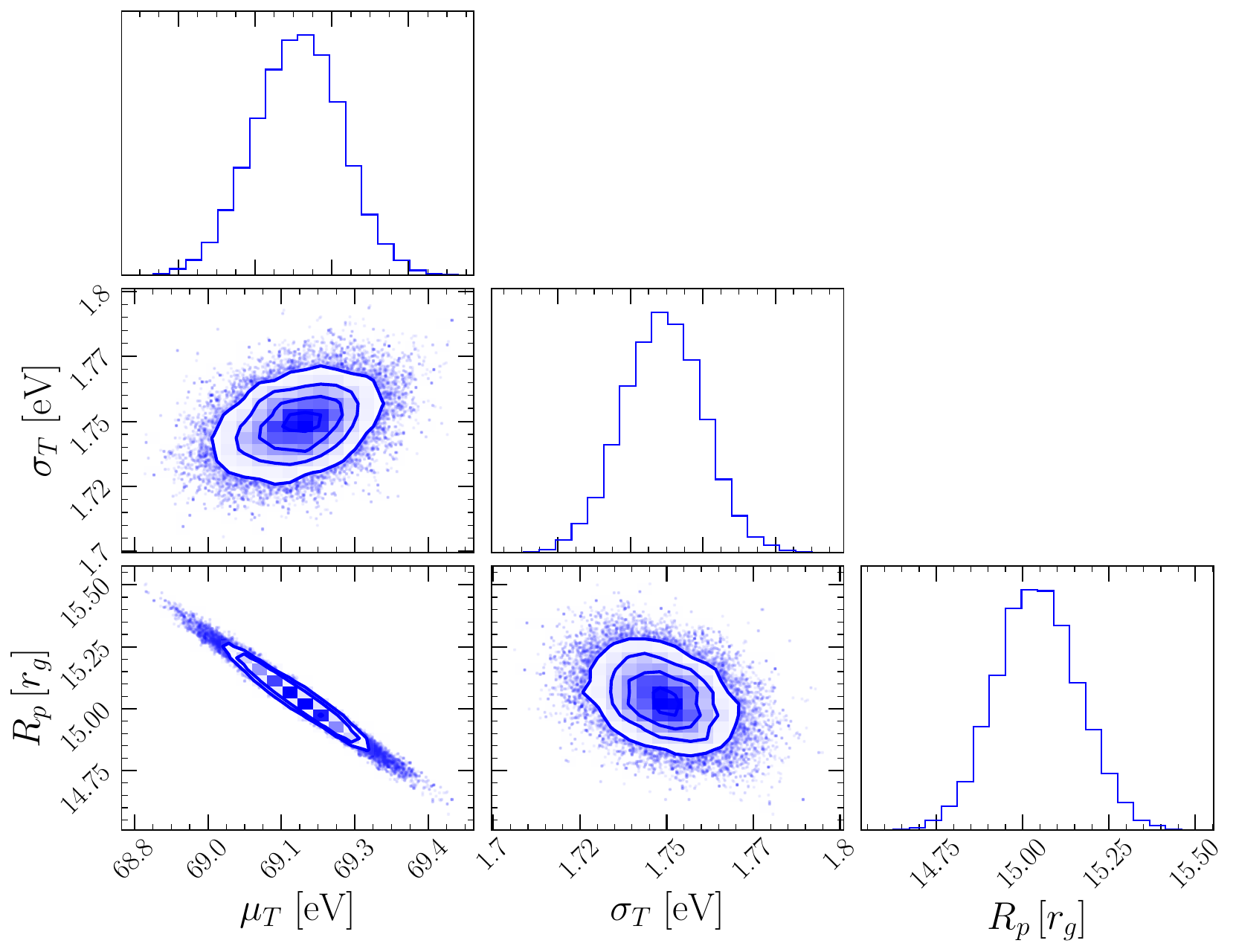}
    \caption{ Upper: Numerical (coloured histograms) and theoretical (purple curves) distributions of the X-ray lightcurves. Each lightcurve was generated with the same disc parameters as in Fig. \ref{fig:fit_to_lc}  (i.e., $M = 10^6 M_\odot, \dot M_0 = 6 \times 10^{21}$ kg/s), except for the coherence length of the stochastic viscosity, which was doubled. The coherence length is anticipated to be of order the scale height of the disc, and so this represents a thicker disc model.  The observing bands were kept identical to Fig. \ref{fig:fit_to_lc}.   The blue (brightest) histogram represents the lowest observing energy, while the red (dimmest) shows the highest.  The purple curves are  fits of the analytical distribution (eq. \ref{xray_dist}) to the data; the fit is excellent. Lower: correlations between the parameters of the X-ray luminosity probability density function, and their posterior distributions. Doubling the coherence length of the stochastic viscosity has approximately doubled the fractional temperature variability $\sigma_T/\mu_T \sim 2.5\%$, and resulted in significantly enhanced luminosity variability.   }
    \label{fig:fit_to_lc_double}
\end{figure}

Increasing the coherence length results in both a higher mean temperature $\mu_T$ and temperature variance $\sigma_T$ being inferred from fits of the analytical distribution (eq. \ref{xray_dist}) to the lightcurves. In Fig. \ref{fig:fit_to_lc_double} we display analogous figures to Figs. \ref{fig:fit_to_lc} and \ref{fig:corner_to_lc}, but now for the lightcurves generated from the $l_c/r = 0.2$ simulation (blue points Fig. \ref{fig:lc_double}). All other parameters of the system are unchanged. We note a few results: the mean temperature has increased slightly (from $k\mu_T \simeq 65$ eV to $k\mu_T \simeq 69$ eV) which results in the larger average X-ray luminosity of the lightcurve (Fig. \ref{fig:lc_double}), while the temperature variance parameter has nearly doubled (from $k\sigma_T \simeq 0.99$ eV to $k\sigma_T \simeq 1.75$ eV). Perhaps surprisingly, the inferred disc size parameter has slightly reduced (from $R_p \simeq 17.8 r_g$ to $R_p \simeq 15r_g$).  These inferred parameters are consistent with the shift in the angle-averaged temperature distribution shown in Fig. \ref{fig:temp_dist_double_length}. 

Importantly, the analytical model of \cite{MummeryBalbus22} is again able to capture the observed variability of each of the different lightcurves with a simple 3-parameter model (we again find $\eta = 5/2$ best describes the data). 

It is not only disc physics which can effect the observed properties of a black hole accretion flows lightcurve however. In the following section we examine how relativistic viewing angle effects may enhance (or suppress) lightcurve variability. 

\section{ Relativistic viewing effects }\label{rel_sec}
The stochastic-viscosity hydrodynamic simulations of \cite{Turner23} are performed using a Newtonian theory of gravity. Naturally however, in real black hole systems there will be important relativistic effects which will modify the properties of observed accretion disc variability. These modifications will come in two distinct forms: the local accretion disc physics will be modified by relativistic gravity \citep[this is particularly relevant for  discs tilted out of the equatorial plane, for example][]{White22}, and in addition there will be relativistic viewing angle effects which also modify the emergent spectrum. These viewing effects include the Doppler-shifting of the emitted photons (owing to the relativistic orbital motion of the fluid elements in the inner disc regions), and the gravitational lensing and red-shifting of the photons over their trajectory to the observer. 

Running numerically expensive full general relativistic magnetohydrodynamic simulations of accretion flows lies well beyond the scope of this paper, and we are not able to examine the effects of strong gravity on the local (disc frame) variability of the accretion flow. Relativistic viewing effects however can be probed, provided that we assume that the simulated discs are evolving around Schwarzschild black holes. While this is of course not a self consistent model (the disc evolves in Newtonian gravity and is then observed in a full relativistic treatment), this modelling is likely to elucidate the key effects of varying the observer's inclination, which as far as the authors are aware is a calculation which has not yet been performed in the literature. We will demonstrate in the following section that these effects can be pronounced, and the fractional variability of the luminosity can double by simply changing the observer's inclination. 

In the following sub-section we will generalise the Newtonian analysis of the previous sections to a fully relativistic treatment, before performing a detailed analysis of the resultant spectra as a function of observer inclination.

\subsection{Relativistic disc spectrum}

\begin{figure*}
    \centering
    \includegraphics[width=.49\linewidth]{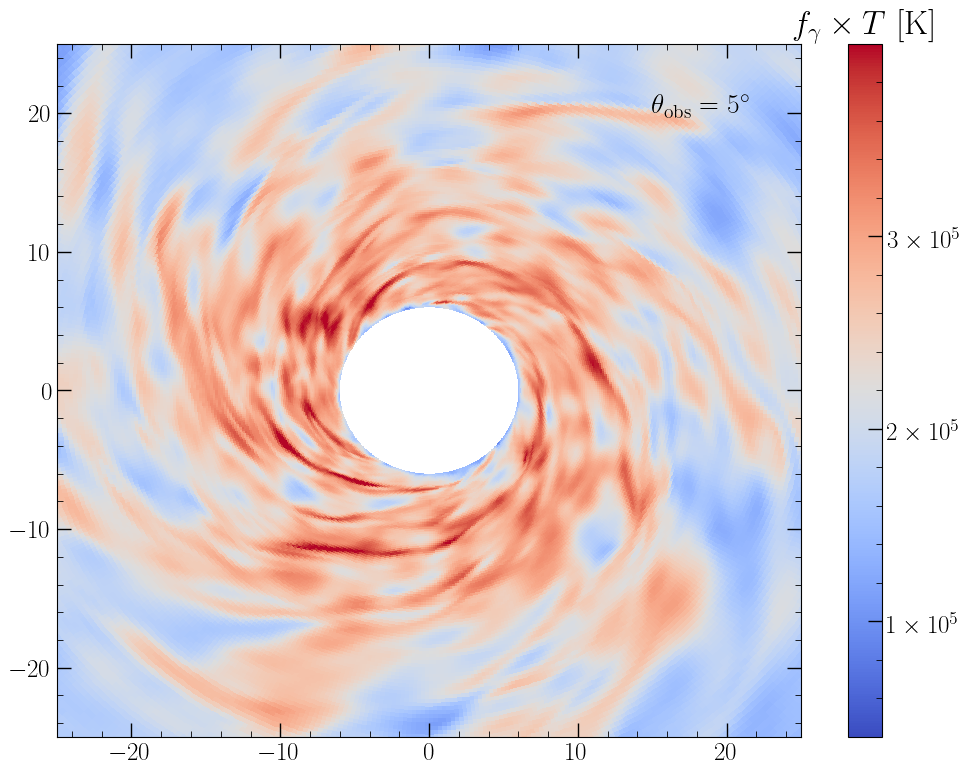}
    \includegraphics[width=.49\linewidth]{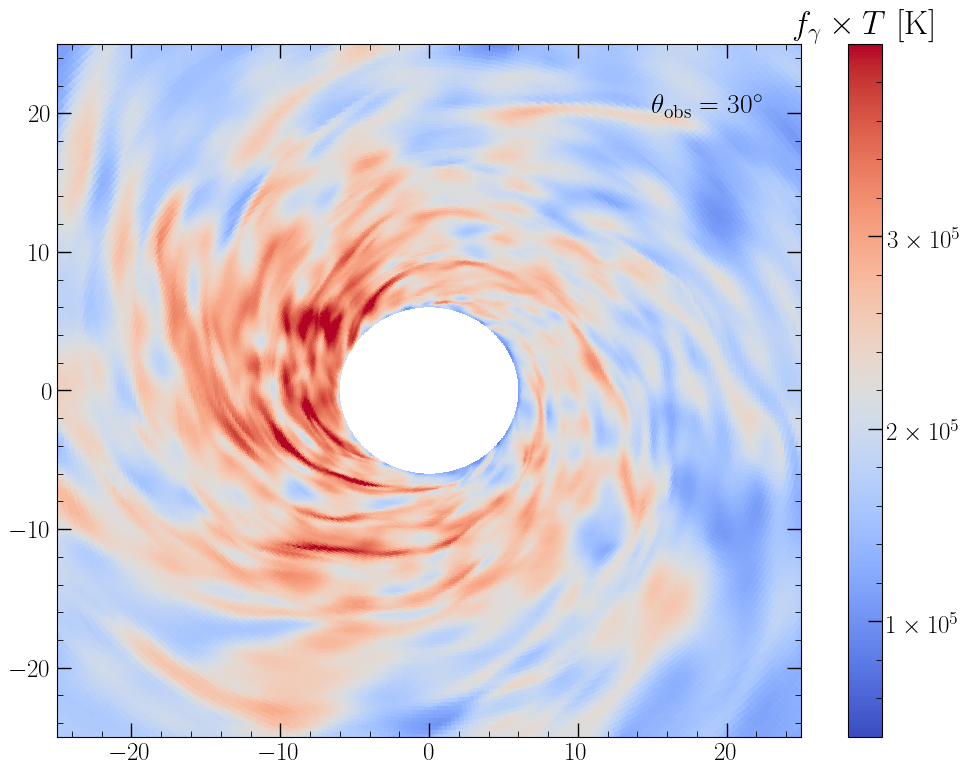}
    \includegraphics[width=.49\linewidth]{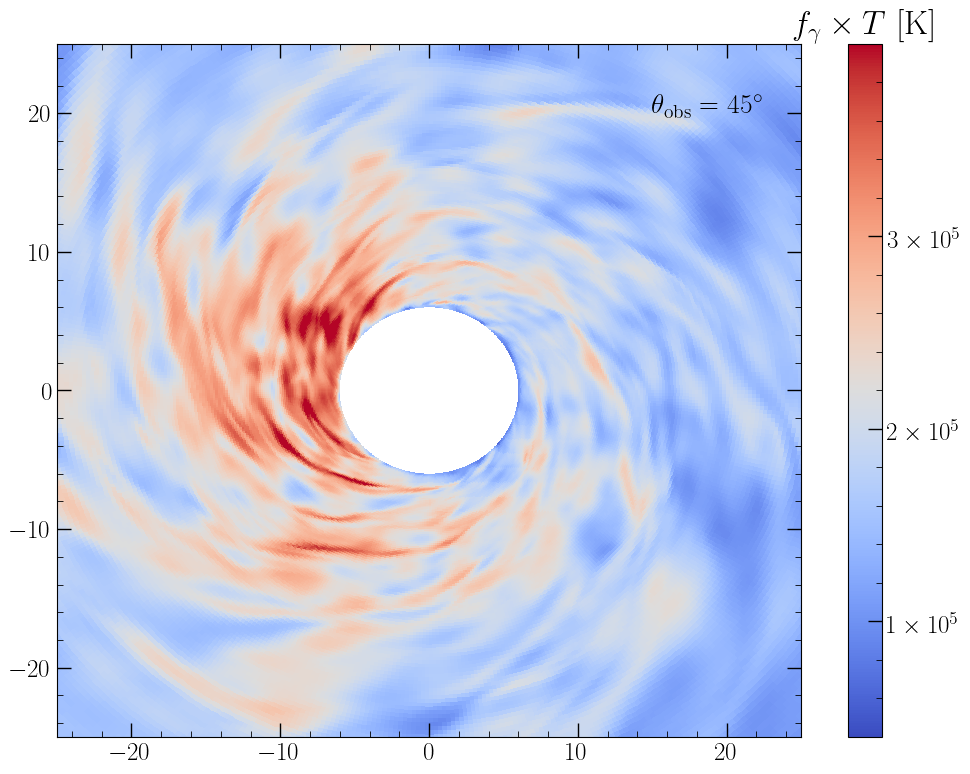}
    \includegraphics[width=.49\linewidth]{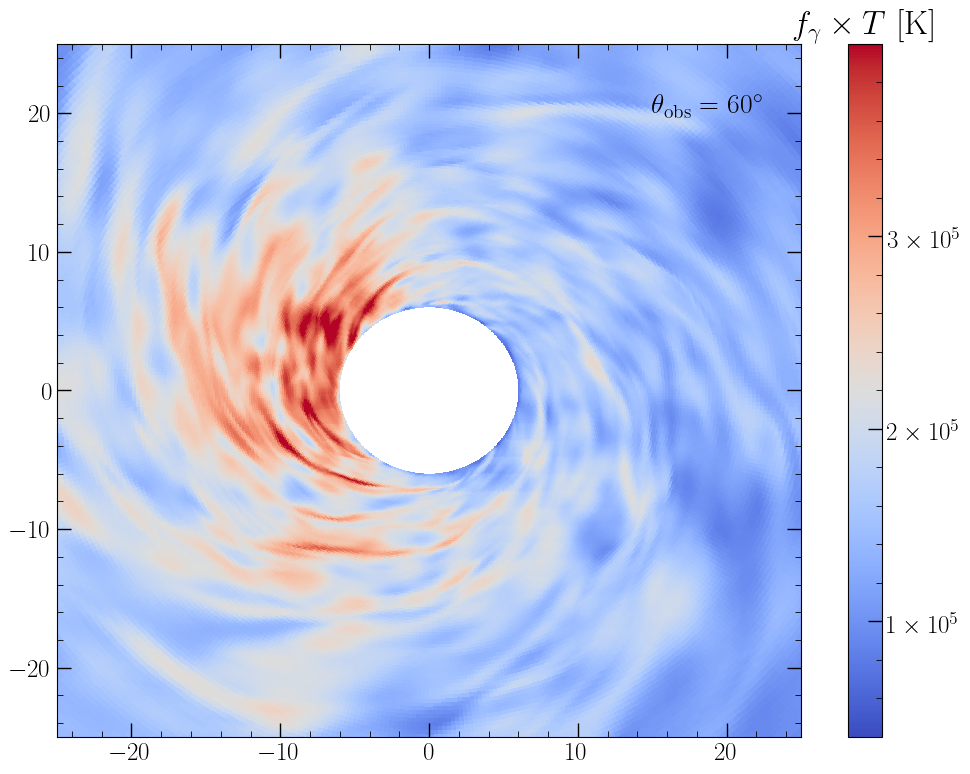}
    \includegraphics[width=.49\linewidth]{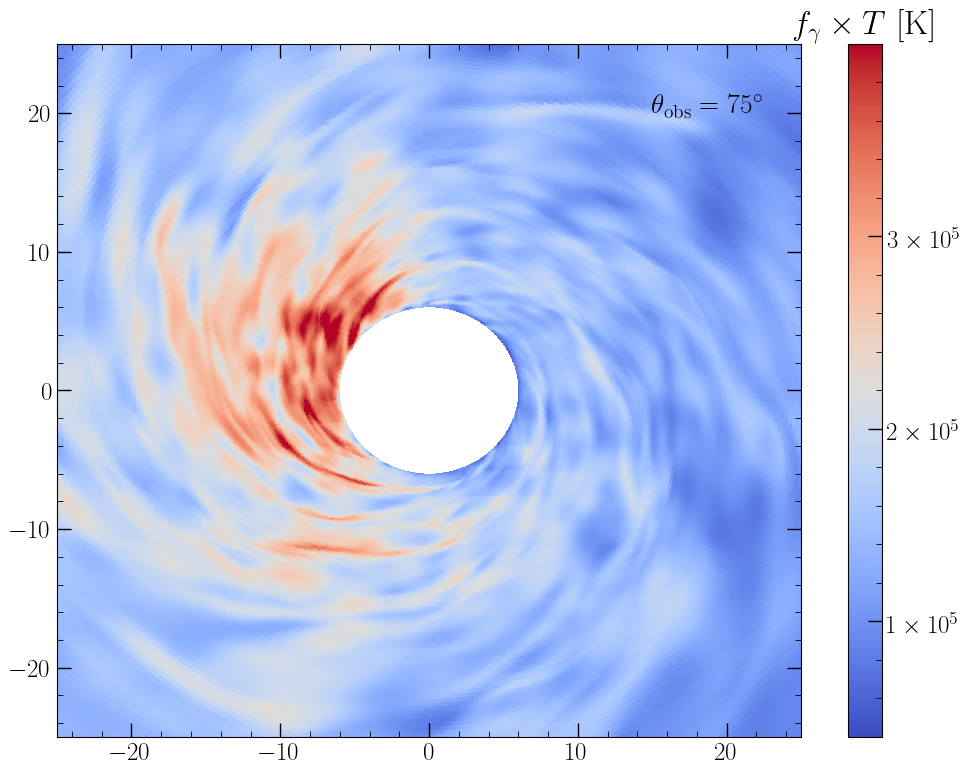}
    \includegraphics[width=.49\linewidth]{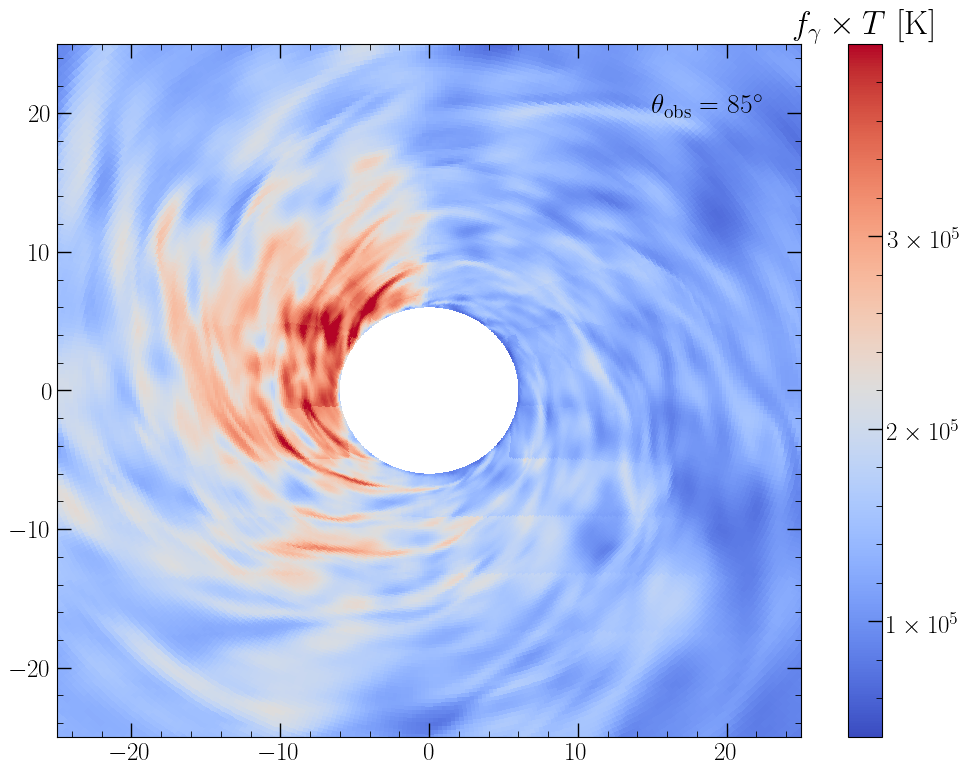}
    \caption{ The ``observed temperature'' of the accretion disc simulations $f_\gamma  T$, projected into the $x-y$ plane of the disc for different inclinations $(\theta_{\rm obs})$ of the distant observer, at a given snapshot in time. The photon ray-tracing calculations assume a Schwarzschild geometry.  The distant observer is orientated at the very bottom of each figure (with $\theta_{\rm obs} = 0^\circ$ being directly above the disc plane), and the disc fluid is rotating in the counter-clockwise direction. The average mass accretion rate for each disc system is different, and was tuned so that the peak observed temperature (and X-ray luminosity) was comparable in each case. The black hole mass is $M = 2 \times 10^6 M_\odot$. Fluid elements on the left hand side of the black hole are observed to be moving towards the distant observer, and get a corresponding Doppler boost to their temperature, while fluid elements on the right are Doppler dimmed. Gravitational red-shifting effects all fluid elements at the same radius equally. Not displayed here are the effects of gravitational lensing, which are taken into account in the luminosity calculations. For larger inclination the observed temperature is peaked in an increasingly small region of the disc, and local variability has a larger effect on the observed luminosity.   }
    \label{fig:viewing_temp}
\end{figure*}

The specific flux density $F_E$ from a source, as observed by a distant observer at rest (subscript ${\rm obs}$), is given by defintion by 
\begin{equation}
F_{E}(E_{\rm obs}) = \iint_{\cal D} I_E(E_{\rm obs} ) \, \text{d}\Omega_{{{\rm obs} }} .
\end{equation} 
Here, $E_{\rm obs} $ is the observed photon energy and $I_E(E_{\rm obs} )$ the specific intensity,  both measured at the location of the distant observer. The integral is to be taken over the entire domain of the disc, denoted ${\cal D}$.   The differential element of solid angle subtended on the observer's sky by a disc area element is $\text{d}\Omega_{{{\rm obs} }}$. 
Since $I_E/ E^3$ is a relativistic invariant \citep[e.g.,][]{MTW}, we may write
\begin{equation}
F_{E}(E_{\rm obs} ) = \iint_{\cal D} f_\gamma^3 I_E(E_{\rm emit}) \, \text{d}\Omega_{{{\rm obs} }},
\end{equation} 
where the photon energy ratio factor $f_\gamma$ is the ratio of $E_{\rm obs} $ to the emitted local rest frame photon energy $E_{\rm emit}$:
\begin{equation}\label{redshift}
f_\gamma(r,\phi) \equiv 
{E_{\rm obs} \over E_{\rm emit}} = {-\left(U^\mu p_\mu\right)_{\rm obs} \over -\left(U^\nu p_\nu\right)_{\rm disc}  } = {1 \over U^t} \left(1 + {p_\phi \over p_t} {U^\phi \over U^t} \right)^{-1} . 
\end{equation}
In this final expression $p_\phi$ and $-p_t$ are the angular momentum and energy of the emitted photon (conveniently constants of motion in the general Kerr metric), $U^t$ and $U^\phi$ are the time and azimuthal 4-velocity components of the orbiting disc fluid. For a Schwarzschild spacetime these components are equal to 
\begin{equation}
    U^t = {1  \over \left(1 - 3r_g/r \right)^{1/2}}, \quad U^\phi =  {\sqrt{GM/r^3}  \over \left(1 - 3r_g/r \right)^{1/2}} ,
\end{equation}
where $r_g = GM/c^2$. 
The spectrum observed from an accretion flow, assuming that it emits thermally in its rest frame, is then equal to 
\begin{equation}\label{rel_spec}
    F_E(E, t) =  \iint_{\cal D} {2  E^3 \over h^3 c^2} { {\rm d}\Omega_{\rm obs}  \over \exp(E/kf_\gamma T) - 1} , 
\end{equation}
where $E$ is the photon energy in the distant observer's rest frame.  The Newtonian expression (eq. \ref{newt_spec}) then simply corresponds to taking the limit $f_\gamma \to 1$, ${\rm d}\Omega_{\rm obs} \to {\cos(i) } \, r \, {\rm d}r \, {\rm d}\phi / D^2$. 

We take the standard approach to computing the relativistic spectral integral (eq. \ref{rel_spec}). Starting with a finely space grid of photon impact parameters (described by cartesian coordinates $b_x, b_y$) in the observer's camera plane, which is inclined at an angle $\theta_{\rm obs}$ from the $z$-axis of the disc,  photon trajectories are traced back until they cross the disc plane $\theta=\pi/2$, at an emission location $(r_e, \phi_e)$. At this location the photon energy shift factor $f_\gamma$ can be computed (eq. \ref{redshift}), and a value for the local temperature $T(r_e, \phi_e, t)$ can be read in from the simulation domain. The differential element of solid angle subtended by each photon is ${\rm d}\Omega_{\rm obs} = {\rm d}b_x {\rm d}b_y/D^2$. The spectral integral (eq. \ref{rel_spec}) can then be computed across the observer's camera plane.  For more details on the algorithm used see Appendix A of \cite{MumBalb20a}.

\begin{figure}
    \centering
    \includegraphics[width=\linewidth]{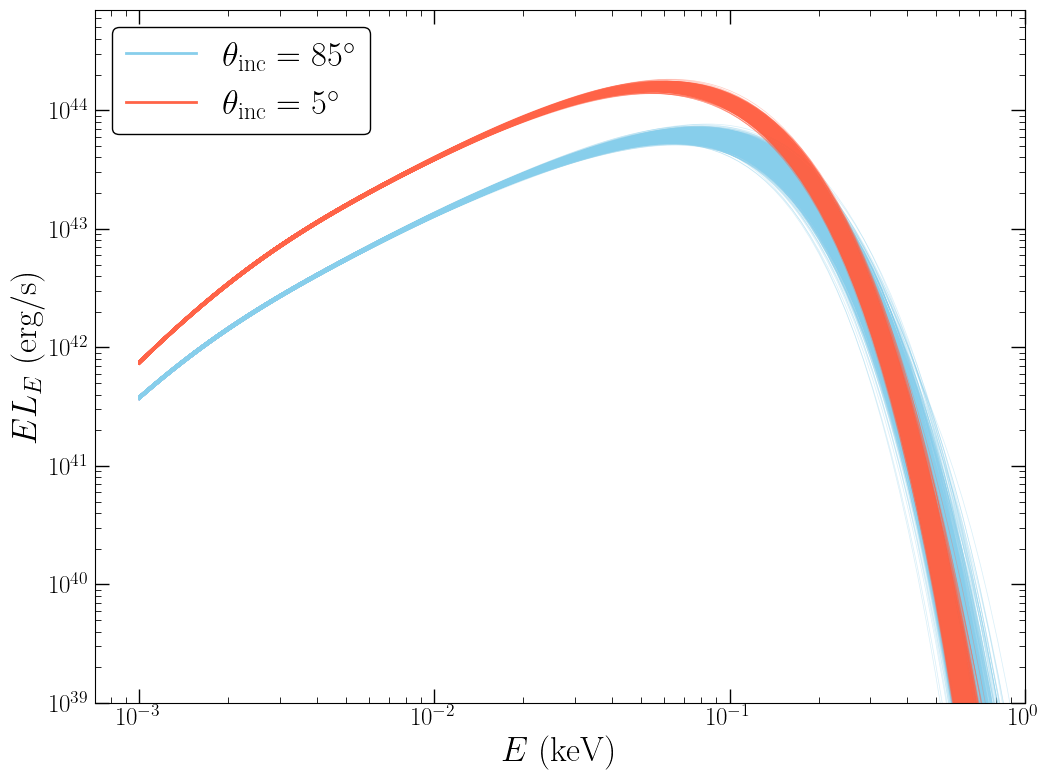}
    \includegraphics[width=\linewidth]{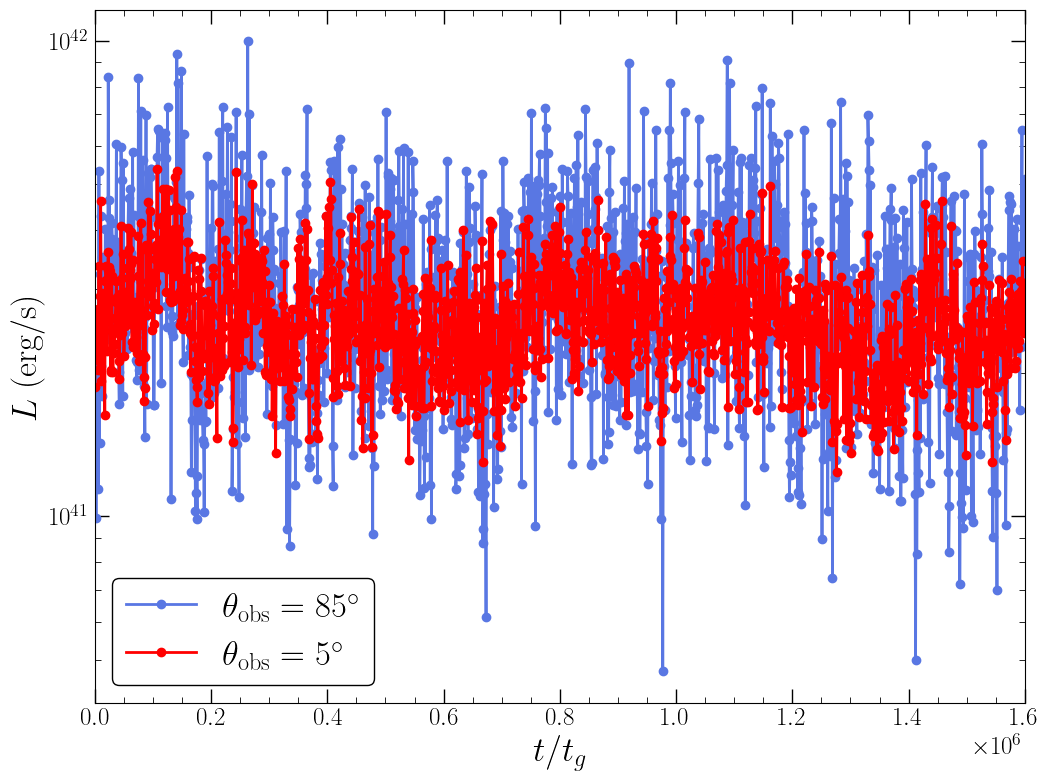}
    \includegraphics[width=\linewidth]{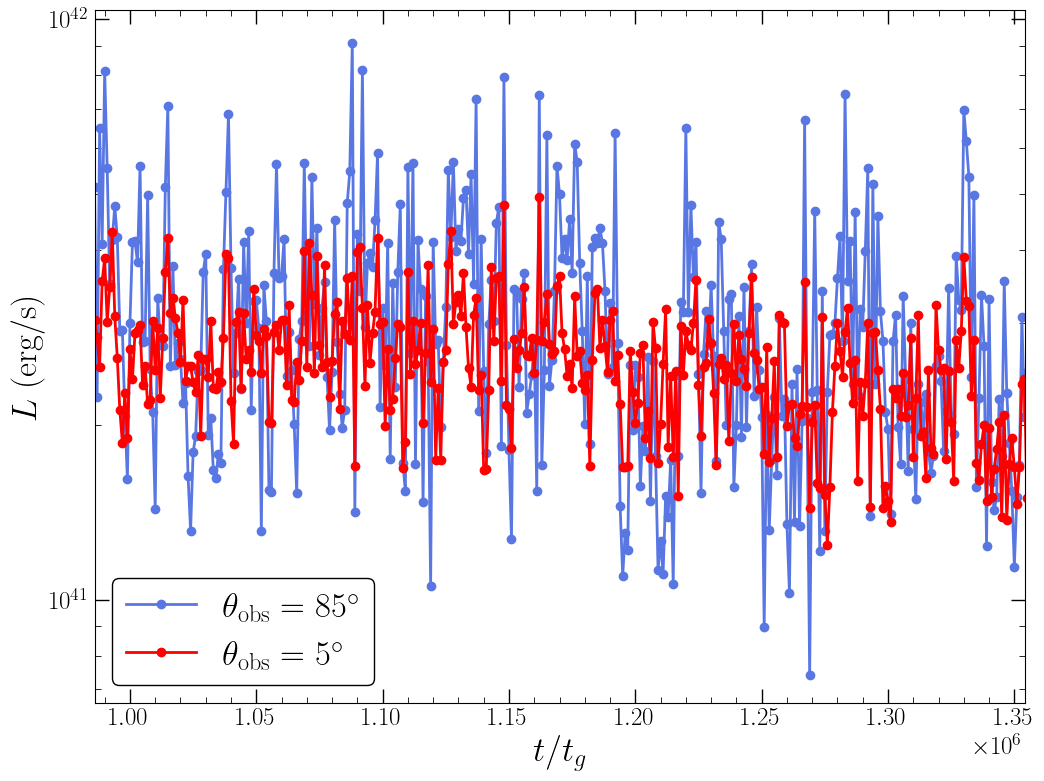}
    \caption{The effect of varying the distant observer inclination on the disc spectra and $0.3-10$ keV lightcurves. In blue we show the lightcurves seen by a near edge-on observer $\theta_{\rm obs} = 85^\circ$, while in red we display the lightcurves seen from a near face-on orientation $\theta_{\rm obs} = 5^\circ$. The mass accretion rate of the two systems are tuned so that they have roughly the same average X-ray luminosity. Owing to Doppler boosting, and the correspondingly smaller disc region contributing to the luminosity of inclined disc systems, the variability of the X-ray lightcurves is substantially increased for near edge-on orientations. Note, as can be seen in the lower zoomed-in panel, that the lightcurves observed at different orientations in a relativistic system need not be precisely correlated with one another, although they often are.  }
    \label{fig:rel_lightcurve}
\end{figure}

\begin{figure*}
    \centering
    \includegraphics[width=.49\linewidth]{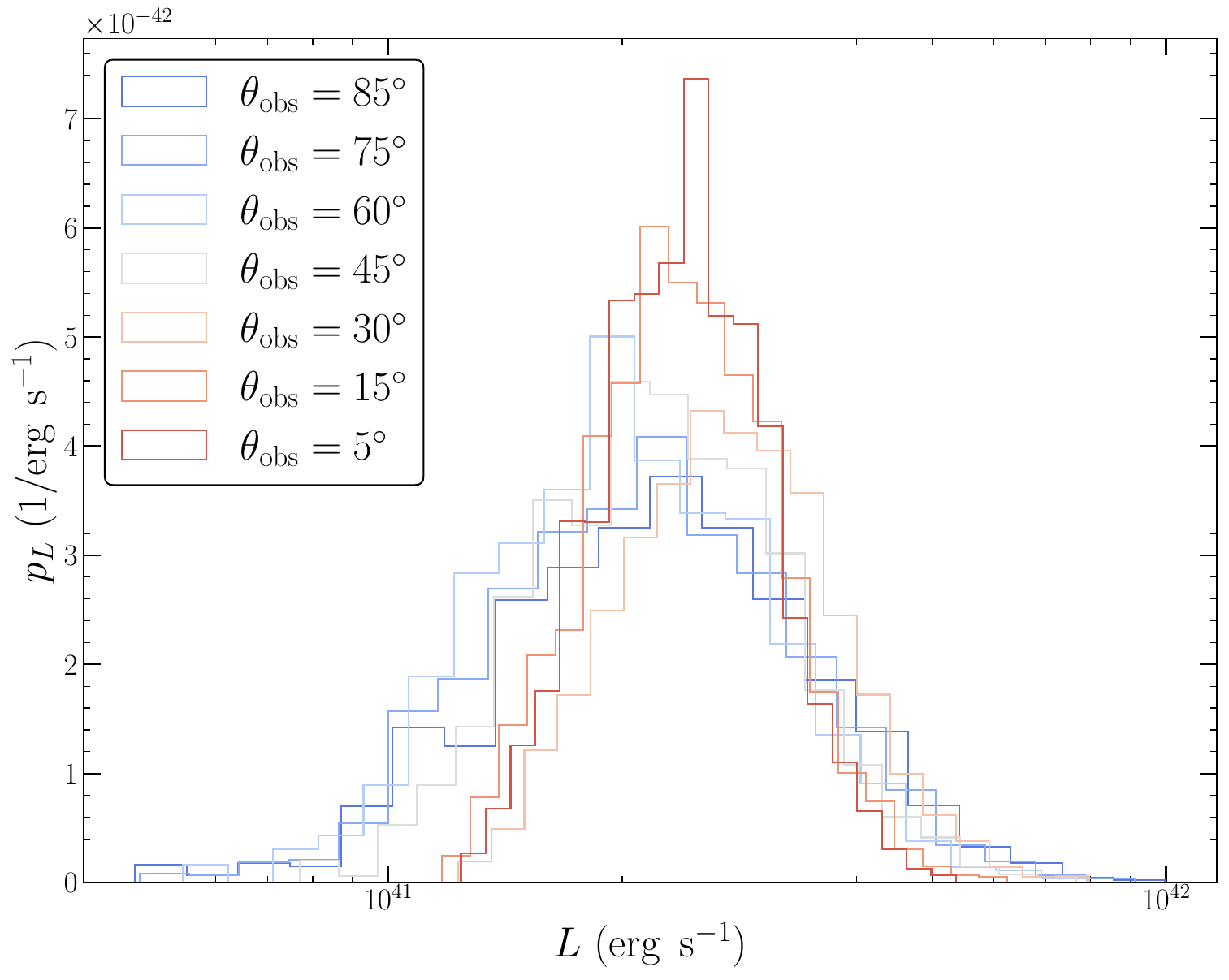}
    \includegraphics[width=.49\linewidth]{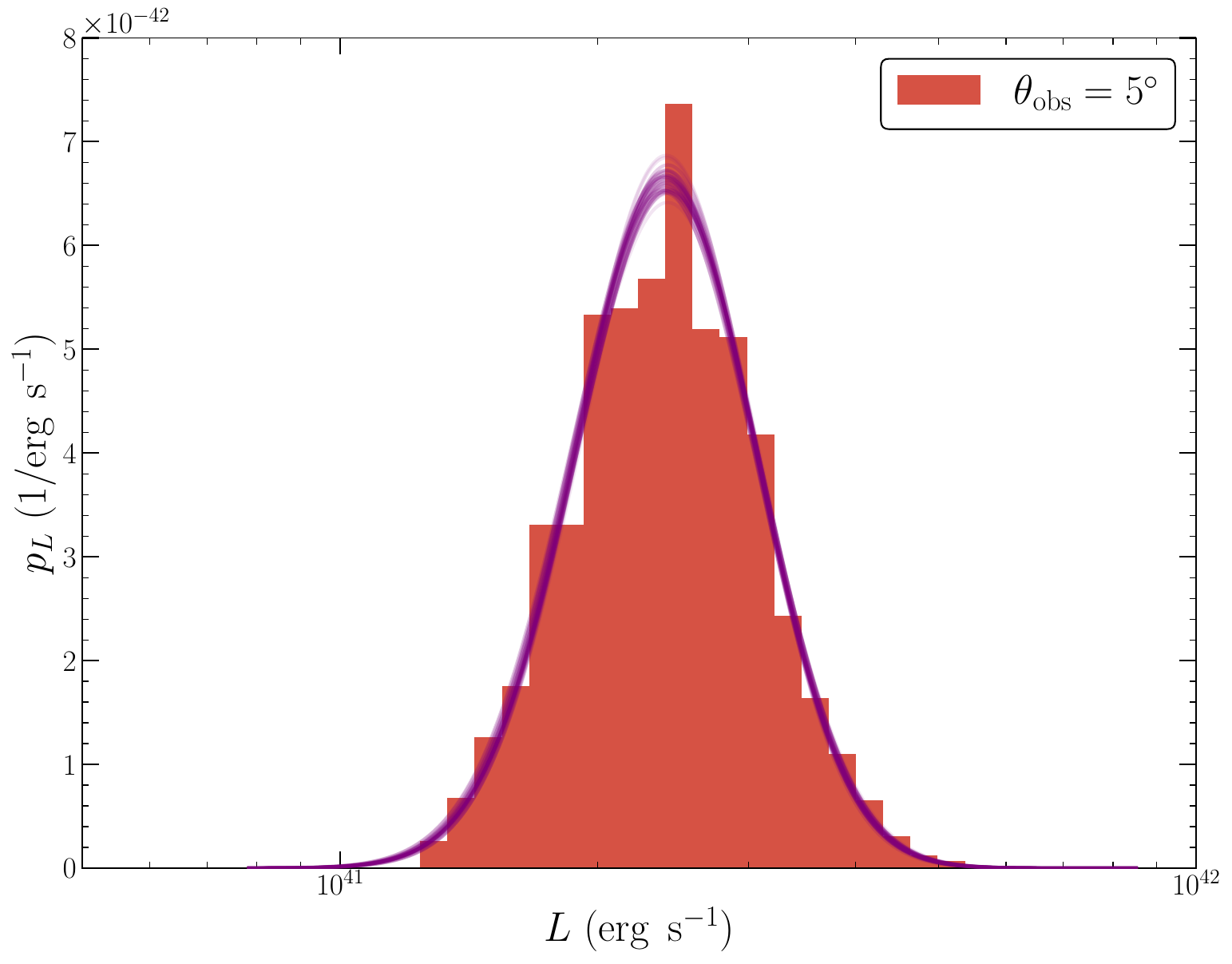}
    \includegraphics[width=.49\linewidth]{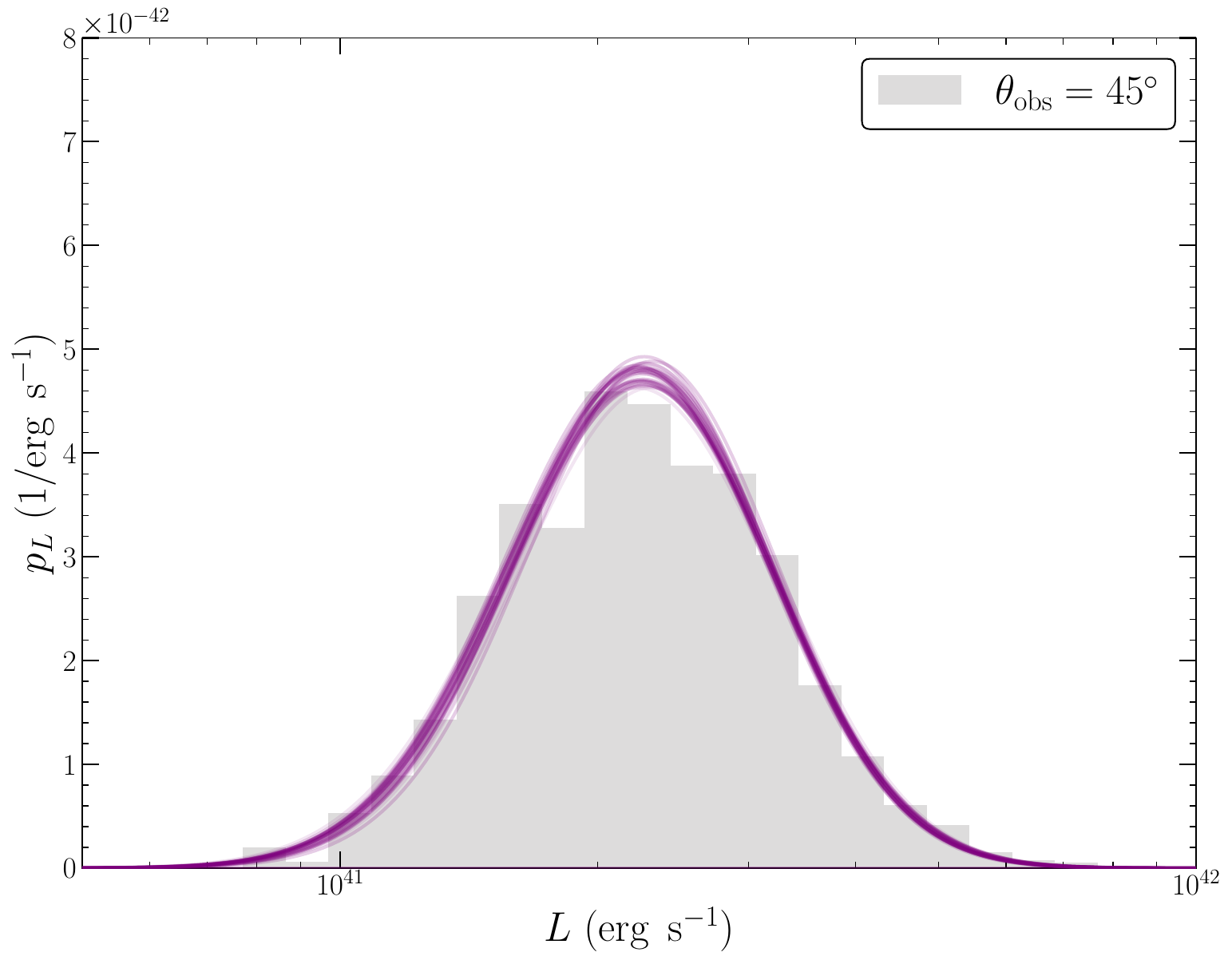}
    \includegraphics[width=.49\linewidth]{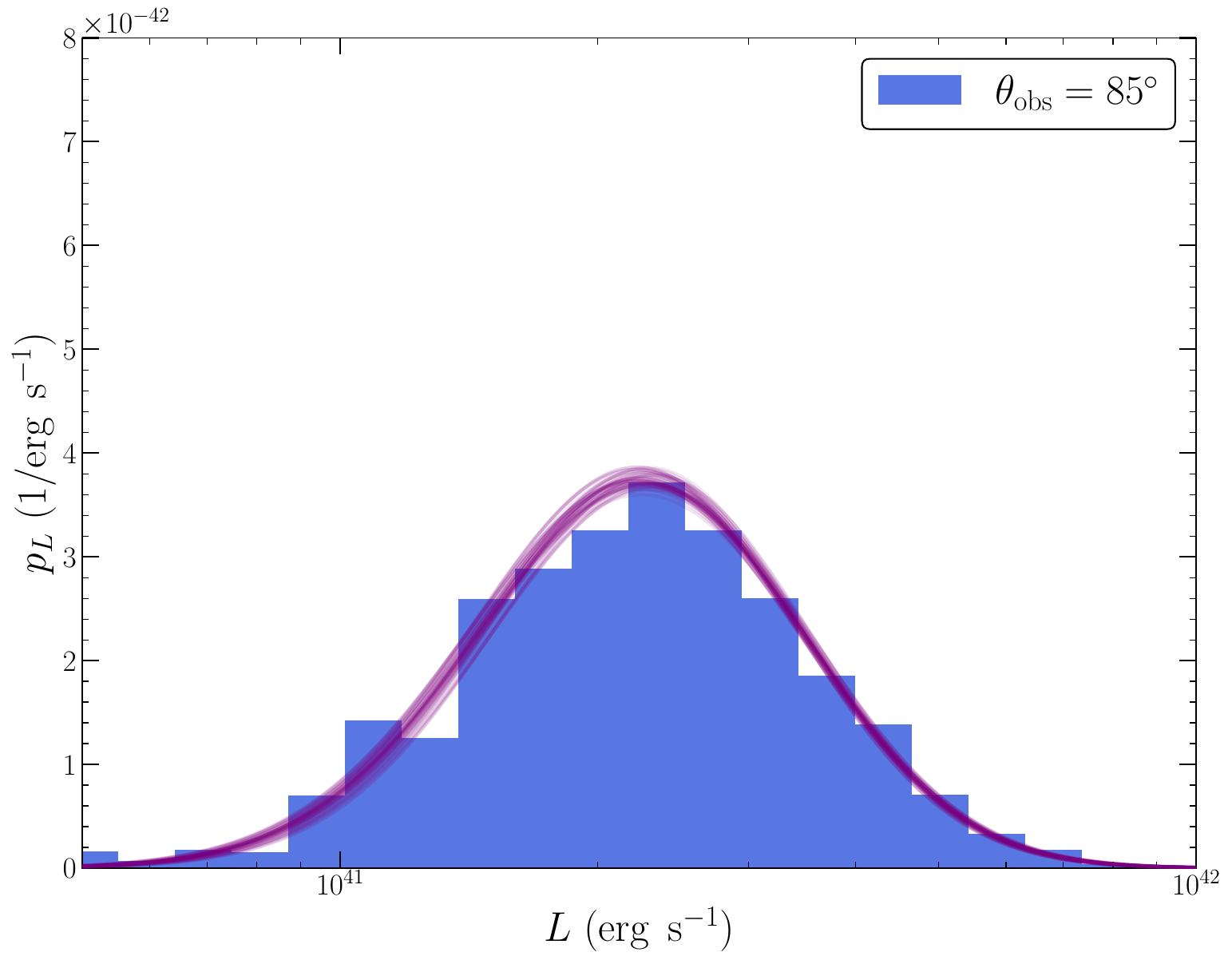}
    \caption{ The X-ray luminosity distributions of relativistic accretion flows observed at increasing inclinations. Upper left: the numerical distributions of the X-ray luminosity for inclinations in the range $\theta_{\rm obs} = 5^\circ - 85^\circ$. The mass accretion rate of the different inclinations were tuned to give roughly the same average X-ray luminosity. The increasing luminosity variance with inclination is clear to see. Upper right,  lower left, and lower right:  individual distributions for three different inclinations, and fits to the observations of the analytical distribution (eq. \ref{xray_dist}; purple curves). The analytical results  well describe the distributions, even with relativistic viewing effects included. The observed temperature variance is a function of inclination, as shown in Fig. \ref{fig:frac_temp_v_inc}.   }
    \label{fig:rel_fits}
\end{figure*}

\subsection{The ``observed temperature'' $f_\gamma T$ }
Equation (\ref{rel_spec}) highlights an interesting property of (thermally emitting) relativistic systems. Namely that the combined effects of gravitational and Doppler energy shifting of photons along their trajectory to the observer result in a simple shifting of the temperature of the emitting material, $T \to f_\gamma T$ (this is precisely analogous to the cosmological red-shifting of the observed temperature of the cosmic microwave background, with $f_\gamma = 1/(1+z)$, for example). 

Therefore, it is the ``observed temperature'', the product of $f_\gamma$ and $T$, which is the relevant quantity for understanding the observed properties of disc lightcurves. Unlike in the cosmological analogy however, for black hole accretion problems this ``observed temperature'' shows a strong observer-inclination dependence, owing to the fact that the inner disc material is orbiting at a substantial fraction of the speed of light, and pronounced Doppler energy-shifts can occur if this motion is directed into the line of sight of the observer. 

By computing energy shift maps $f_\gamma(r, \phi)$ of photons emitted from a disc location $(r_e, \phi_e)$ and ultimately observed by a distant observer inclined at an angle $\theta_{\rm obs}$ from the $z$-axis of the disc plane, it is possible to examine the two dimensional temperature profiles of the \cite{Turner23} disc models, {\it as observed by a distant observer}.  These two dimensional maps are shown in Fig. \ref{fig:viewing_temp}, we remind the reader that these maps were produced assuming a Schwarzschild spacetime. 

In Fig. \ref{fig:viewing_temp} we plot two dimensional profiles of the ``observed temperature'' of the accretion disc simulation, defined as the product $f_\gamma  T$, projected into the $x-y$ plane of the disc for different inclinations $(\theta_{\rm obs})$ of the distant observer (displayed in the upper right corner of each panel). Each snapshot is taken at the same instant in time, only the viewing angle is different between panels. The distant observer is located at the very bottom of each figure (with $\theta_{\rm obs} = 0^\circ$ being the orientation directly above the disc plane), and the disc fluid is rotating in the disc plane in the counter-clockwise direction. The average mass accretion rate for each disc system is different, and was tuned so that the peak observed temperature (and X-ray luminosity) was comparable in each case. The black hole mass is $M = 2 \times 10^6 M_\odot$, and for the remainder of this section we use the simulation with $l_c/r = 0.2$ (i.e. as seen in the bottom panel of Fig. \ref{fig:disc_phys}, which shows the temperature snapshot in a Newtonian theory), to maximise the observed variability. 

The effects of relativity introduce a clear asymmetry into the observed temperature profiles for inclined systems $\theta_{\rm obs} \gtrsim 45^\circ$. The interpretation and physical cause of this result is simple, fluid elements on the left hand side of the black hole (in this Figure) are moving with a significant component of their velocity pointing towards the distant observer, and get a corresponding Doppler boost to their temperature $f_\gamma > 1$, while fluid elements on the right are moving away from the observer and are therefore Doppler dimmed $f_\gamma < 1$. Gravitational red-shifting effects all fluid elements at the same radius equally, and has the net effect of reducing the observed temperature at all radii, but does not introduce an asymmetry to the profile. 

Not displayed in this Figure are the effects of gravitational lensing (i.e., the bending of light rays which pass close to the central black hole). These effects are however taken into account in the full luminosity calculations. For larger inclination the disc temperature is peaked in an increasingly small region of the disc, with an observer in the Wien-tail only ``seeing'' a small wedge of the disc for $\theta_{\rm obs} = 85^\circ$, while a face-on observer would ``see'' contributions from the full $2\pi$ azimuthal range of the disc.

\begin{figure}
    \centering
    \includegraphics[width=\linewidth]{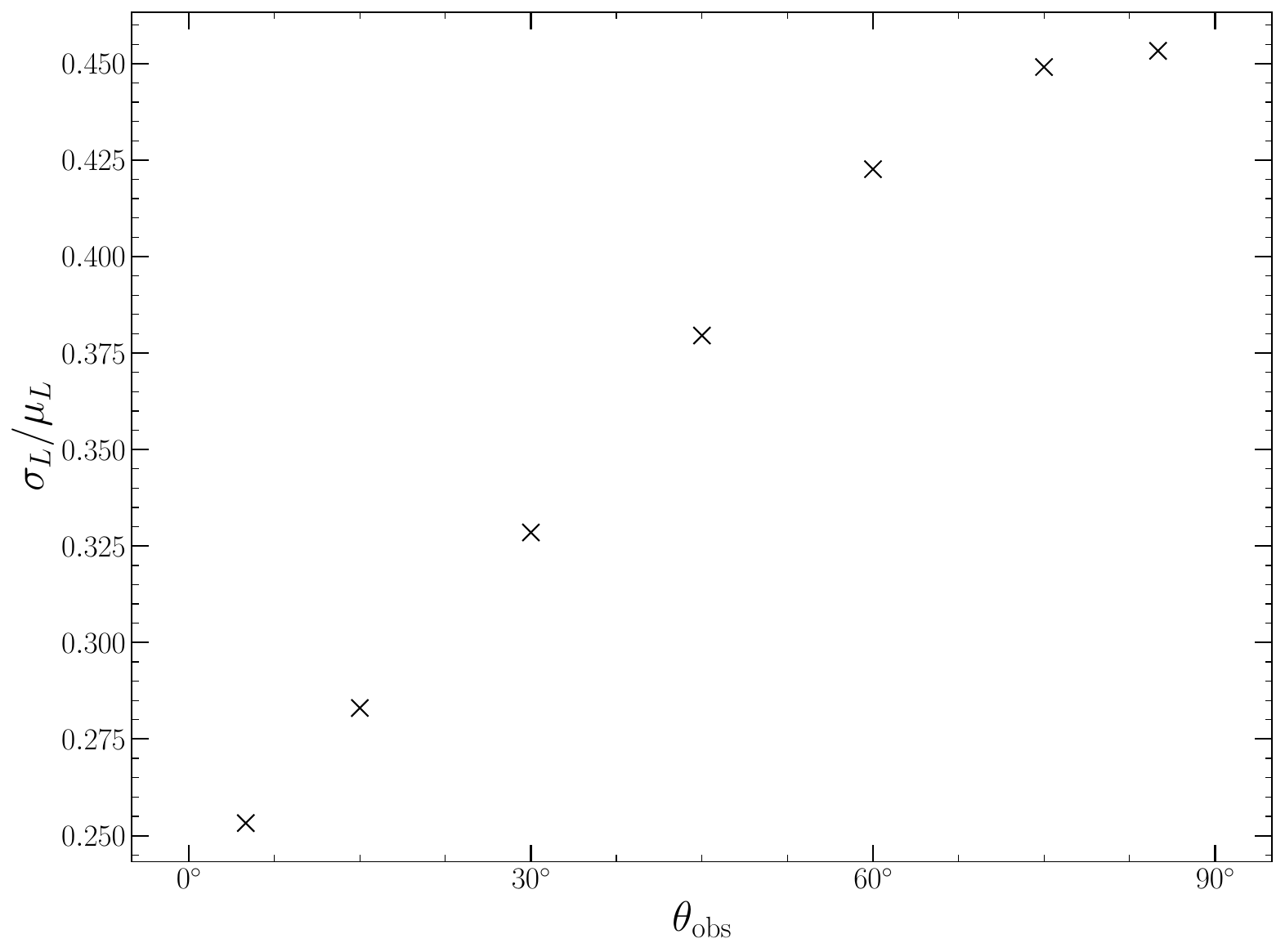}
    \includegraphics[width=\linewidth]{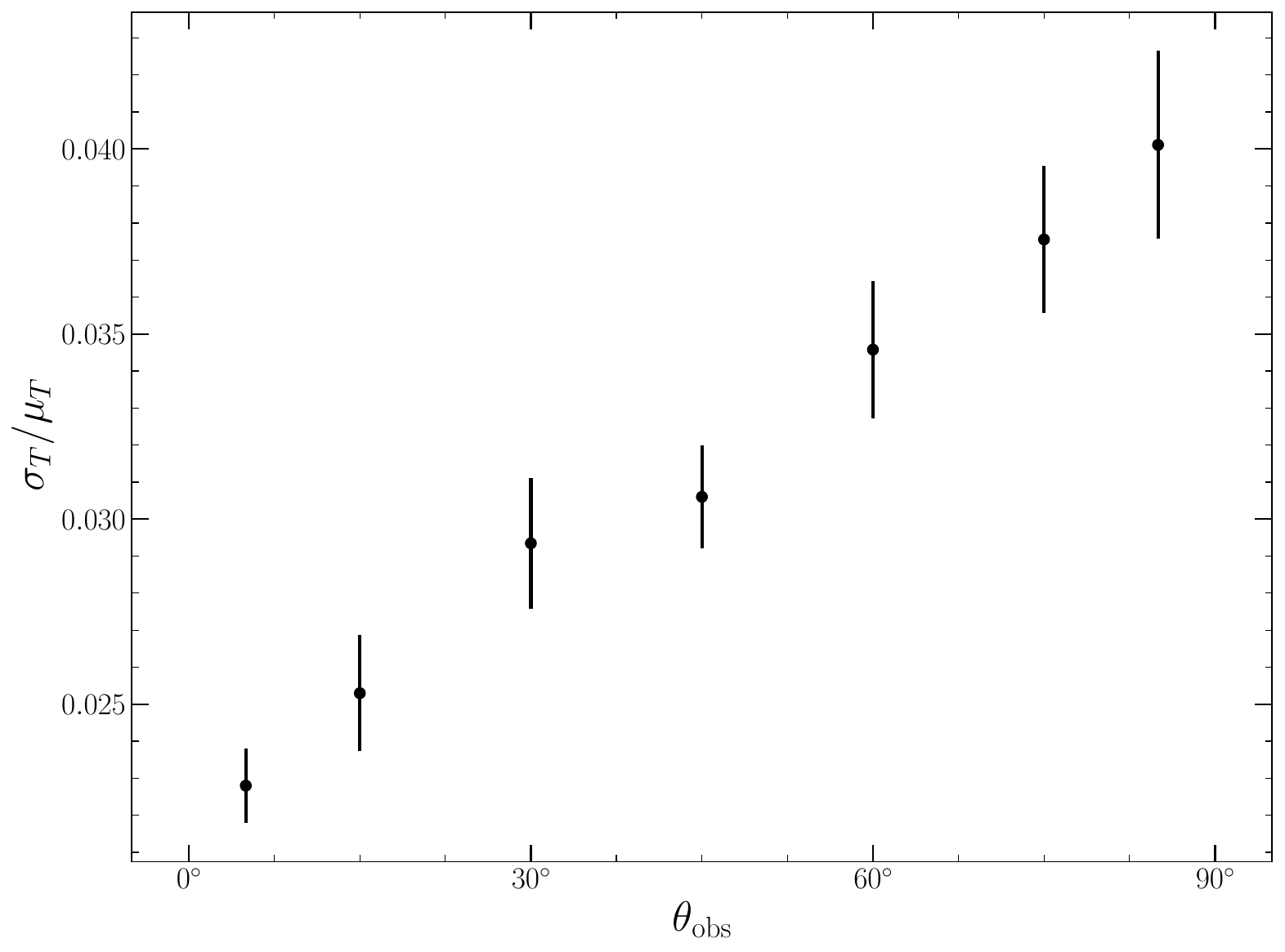}
    \caption{Upper: the fractional X-ray luminosity variability as a function of observer inclination. Lower: the inferred fractional temperature variability, from fits of the probability distribution (eq. \ref{xray_dist}) to the lightcurves, also as a function of observer inclination. Error bars denote the standard deviation of the fit posteriors. The fractional temperature variability roughly doubles with inclination, as does the fractional luminosity variance.    }
    \label{fig:frac_temp_v_inc}
\end{figure}

\subsection{ Lightcurve variability as a function of inclination }

In a relativistic system the observer's inclination angle has a large effect on the observed variability of a turbulent disc system.  This is highlighted in Fig. \ref{fig:rel_lightcurve}. In Fig. \ref{fig:rel_lightcurve} we display the full lightcurves generated from the 1600 snapshots of the disc temperature recorded in the simulations of \cite{Turner23}, as observed at an inclination of $\theta_{\rm obs} = 5^\circ$ (red curve) and $\theta_{\rm obs} = 85^\circ$ (blue curve); these inclinations correspond to the upper left and lower right disc temperature snapshots of Fig. \ref{fig:viewing_temp}. The black hole mass of both systems was $M = 2 \times 10^6 M_\odot$, while the average accretion rate $\dot M_0$ of the two systems were tweaked so that they had roughly the same average X-ray ($0.3-10$ keV) luminosity. The accretion rate had to be reduced by a factor of $\sim 3$ over the range $\theta_{\rm obs} = 5^\circ - 85^\circ$, with more edge-on discs brighter in the Wien-tail for fixed accretion rate (owing to Doppler boosting, which more than counteracts the usual reduction in the luminosity due to the $\cos i$ term in eq. \ref{newt_spec}; see the upper panel for more details). 

For large (near edge-on) inclination angles the disc regions with the largest observed temperatures shrinks to a small region with fluid elements moving towards the observer, and only this small region of the disc contributes to the observed Wien-tail luminosity. This means that observations of the Wien-tail are sensitive to an ever smaller region of the disc, and are therefore increasingly sensitive to the local physics of the disc variability. This results in a substantially larger fractional disc variability in the case of large inclinations (Fig. \ref{fig:rel_lightcurve}). Physically, variations on the order of a few coherence lengths $l_c$ show substantial variability amplitudes; when the flux is made up of an average over the entire azimuthal disc structure these local fluctuations are somewhat averaged out. When only a small wedge of the disc contributes to the flux, these variations show an enhanced effect.

An interesting result to note is that, as can be seen in the lower zoomed-in panel,  the lightcurves observed at different orientations in a relativistic system need not be precisely correlated with one another, although they often are somewhat correlated. This highlights the point that different disc regions (more precisely different fractions of the disc angular extent) are being ``seen'' by different observers.

The model of \cite{MummeryBalbus22} (eq. \ref{xray_dist}) still describes well the observed distributions of the disc luminosity, even with relativistic viewing angle dependence included. This is a result  of the ``observed temperature'' property of thermal relativistic systems, meaning that the gross functional form of the Wien-tail luminosity is unchanged by including relativistic effects. In Fig. \ref{fig:rel_fits} we display the numerical distributions of the observed $0.3-10$ keV X-ray luminosities for each of the viewing angles displayed in Fig. \ref{fig:viewing_temp}. The average mass accretion rate was tweaked in each case to keep the average luminosities roughly the same. It is clear to see that the fractional luminosity variability $\sigma_L/\mu_L$ increases with increasing inclination (upper left panel). In the upper right, lower left and lower right panels we show fits of the analytical probability density function (eq. \ref{xray_dist}) to each of the numerical distributions, for a range of observer inclinations. The fit in each case is good, and the different distributions result simply from differing inferred disc temperature parameters, chiefly the differing temperature variance $\sigma_T$ as a function of inclination.

In Fig. \ref{fig:frac_temp_v_inc} we display both the fractional luminosity (upper) and disc temperature (lower) variability, as a function of observer inclination.  Error bars on the lower plot denote the standard deviation of the fit posteriors.  The fractional temperature variability $\sigma_T/\mu_T$ roughly doubles as the inclination is varied form $5^\circ$ to $85^\circ$, as does the fractional luminosity variance which peaks at $\sigma_L/\mu_L \sim 0.45$, which represents extreme variability. 

\section{Discussion -- tidal disruption events as a probe of ISCO-scale disc turbulence }\label{discussion}
In this paper we have shown, provided that the evolution of real accretion flows is well approximated by a stochastic-viscosity hydrodynamic theory (as in the standard theory of propagating fluctuations), that tidal disruption events are ideal targets to study stochastic variability in accretion flows, as they may be observed to show extreme variability in the standard $0.3-10$ keV X-ray bands. The reason that tidal disruption events represent particularly interesting sources from the perspective of variability studies is multi-faceted, and we list the principle reasons below. 

Firstly, and importantly, tidal disruption event discs typically have peak disc temperatures at the scale of $kT_p \lesssim 100$ eV \citep[e.g.,][]{Bright18, Wevers19b, Saxton21, Mummery_Wevers_23}, a factor at least 3 lower than the typical lower bandpass energy of X-ray telescopes $E_l = 300$ eV. This means that tidal disruption events are observed robustly in the Wien-tail of the disc spectrum, where any variability is exponentially amplified \citep{MummeryBalbus22}. As we have demonstrated in this paper, relatively  minor variability in the angle-averaged disc temperature profile (Fig. \ref{fig:temp-dists}) can translate into factor unity variability in Wien-tail light curves (Fig. \ref{fig:fit_to_lc}), up to scales even as high as an order of magnitude for inclined systems (Fig. \ref{fig:rel_lightcurve}). 

Secondly, tidal disruption event discs are expected to be somewhat more compact and relatively thicker ($H/R \gtrsim 0.1$) than the analogous accretion states of X-ray binary and AGN discs, owing to the relatively high accretion rates of post-disruption debris fallback \cite{Rees88}.  As the coherence length of typical turbulent eddies is expected to be of order the scale height of the disc $l_c \sim H$, this means that tidal disruption event disc turbulence may well be coherent over larger scales. Equivalently, there are fewer independent disc regions near to the ISCO ($N \sim 2\pi r_I / l_c$), and the intrinsic variability of these systems is larger (Figs. \ref{fig:temp_dist_double_length}, \ref{fig:lc_double}, \ref{fig:fit_to_lc_double}). 

In addition, the large black hole masses of typical tidal disruption event hosts $M_\bullet \sim 10^6-10^8 M_\odot$ makes the orbital timescale of the innermost disc regions $t_{\rm orb} = \sqrt{r^3/GM_\bullet} \sim {\cal O}(1-10)$ hours. This is ideal for probing the short timescale variability with dedicated follow up X-ray observations on day--week timescales. 

Finally, tidal disruption event X-ray spectra typically do not show pronounced power-law components which would otherwise mask the Wien-tail. This is in sharp contrast to both X-ray binary and AGN spectra. As it is unclear how variability of a Comptonized component tracks the intrinsic variability of the accretion flow itself, observations of the soft disc-dominated component appears the best route probing intrinsic disc turbulence. 

For these reasons, we recommend dedicated X-ray observational campaigns of future tidal disruption events to aid in our understanding of disc turbulence. Conversely, this can be phrased as a prediction: we expect that future short-timescale resolved analysis of X-ray bright TDEs will detect $\sim$ order of magnitude X-ray flares on $\sim$ hour-to-day timescales. 

{One potential complication in probing the physics of disc turbulence with tidal disruption events is the possibility that these sources may be accreting at super-Eddington rates at early times, where it is less clear to what degree the predictions of the theory of propagating fluctuations robustly hold. A simple ``fall-back rate'' calculation implies super-Eddington accretion in a typical tidal disruption event for the first $\sim$ year of its evolution \citep{Rees88}, and it has been suggested that at these high accretion rates viewing angle effects (chiefly whether or not the observer's line of sight passes through a thick super-Eddington torus) may dominate the observed properties of the source   \citep{Dai18}.  }

{That being said, while disc accretion at super-Eddington rates must to some degree be more complex than the thinner discs considered in this work, the main results presented here are based on simple physical principles which should be universal, and apply  more broadly than to just thin discs. It seems extremely likely that all disc accretion (super-Eddington or otherwise) will remain a turbulent process, and therefore that all disc emission profiles will be stochastic to some degree. The exponential amplification of these stochastic features by the Wien tail is dependent  only on the thermal nature of this emission and the observer's bandpass energy, and not on the broader physical nature of the emitter itself. It seems likely that even at early times therefore large amplitude fluctuations should be observed in tidal disruption event X-ray light curves, while at later times these systems will approach the precise theory developed here \citep[tidal disruption events are routinely X-ray bright for $\sim 1000$ days, e.g.,][]{Bright18}.  }

\subsection{Real tidal disruption events}
Given this prediction of the theory of propagating fluctuations, it is interesting to look at the X-ray lightcurve properties of known tidal disruption events to see if any hints of the behaviour described here have been previously observed. 

Understanding the properties of TDE X-ray light curve fluctuations will only grow in importance as the sample size of X-ray bright TDEs increases in the coming years. Indeed, the current sample of X-ray bright TDEs already appear to show significant X-ray variability, but due to sampling issues in follow-up observations (for example, non uniform observing cadences and big gaps between observations), such behaviour is currently poorly characterised. Despite this, examples of different types of behaviour already observed include rapid ($\sim$weeks) flaring/dipping episodes \citep{Saxton12, vanVelzen20}, sudden \citep{Kajava20} or smooth \citep{Gezari17}  X-ray brightening with significant delays with respect to the UV/optical peak. While a full analysis of the lightcurve variability of tidal disruption events is beyond the scope of this work, and is postponed for future study, we do highlight one particularly interesting source: the tidal disruption event AT2019azh.

\begin{figure}
    \centering
    \includegraphics[width=\linewidth]{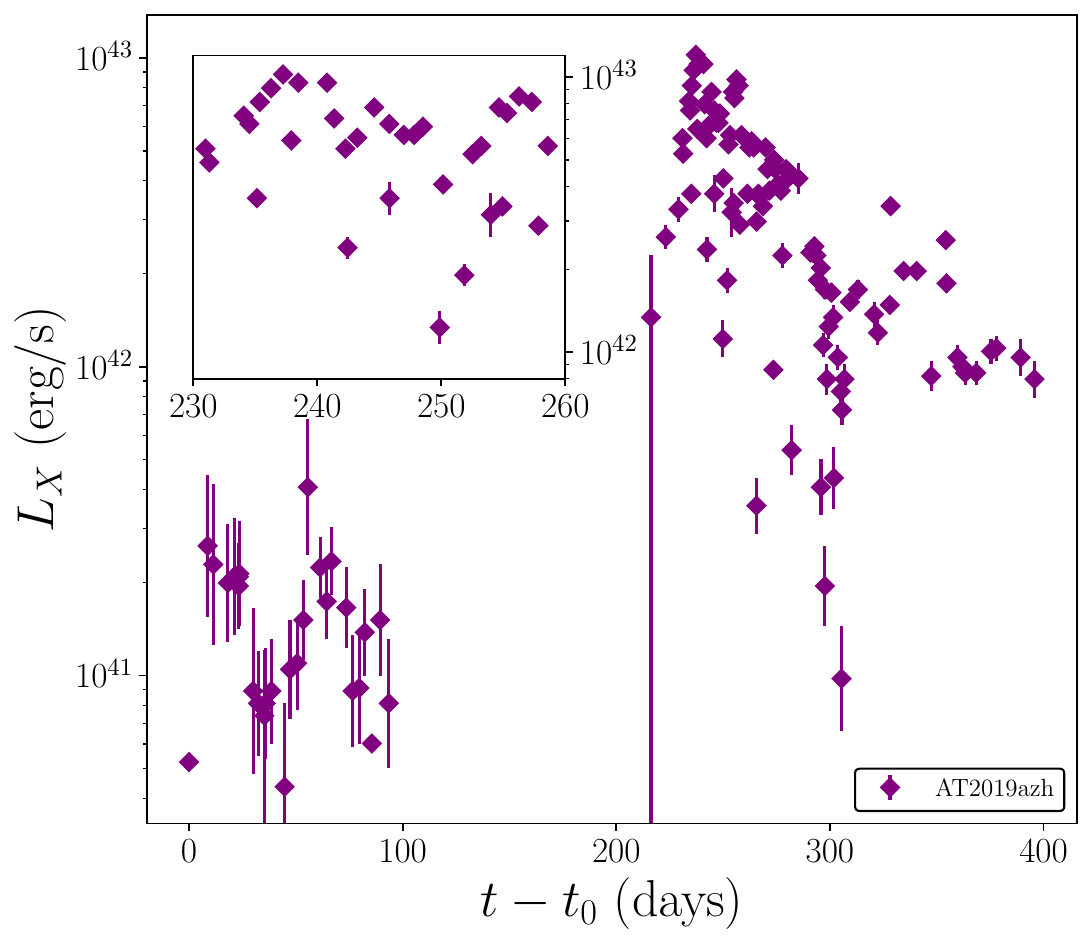}
    \caption{ The X-ray ($0.3-10$ keV) lightcurve of the tidal disruption event AT2019azh. This source shows clear large-amplitude stochastic variability on timescales as short as the observing cadence (see inset for a zoom-in of one month of data around 230 days after first detection). The X-ray spectrum of this source is dominated by a cool $kT \sim 50$ eV thermal component, as expected from the models put forward in this paper. The error bars on the data points may be smaller than the marker size.  }
    \label{fig:real-tdes}
\end{figure}

In Fig. \ref{fig:real-tdes} we display the X-ray lightcurve of AT2019azh \citep{Hinkle21, Goodwin22}, which is an X-ray bright tidal disruption event with a cool ($k T \sim 50$ eV) but thermal X-ray spectrum \citep{Hinkle21, Mummery_Wevers_23}. In the latter ($t > 200$ days since first detection) stages of it's X-ray lightcurve evolution AT2019azh shows extreme X-ray variability, on timescales at least as short as the observational timescale. During this phase of evolution the spectrum was very soft, and dominated entirely by a thermal component \citep{Hinkle21, Mummery_Wevers_23}. There is also a robust late time detection of an optical plateau in the AT2019azh lightcurve, indicating the presence of an accretion disc \citep{Mummery_et_al_24}. 

It may well be the case that AT2019azh represents an observational ``detection'' of the theory of propagating fluctuations, in its simplest form. There is clearly no periodic structure in the X-ray variability of AT2019azh (ruling out, for example, a repeating tidal disruption event), which appears to be stochastic. 
The amplitude of variability shown by AT2019azh is extremely large, as is expected by the exponential enhancement of turbulent temperature variability. 

In a follow up work we will aim to quantitatively analyse AT2019azh (and other sources like it) with the numerical and analytical models presented in this paper. 

\section{ Conclusions }\label{conclusion}
In this paper we have analysed the properties of accretion disc variability (driven by fluctuations in the angular momentum transport) as observed at high photon energies (i.e., photon energies larger than the peak disc temperature). We have shown that small amplitude temperature fluctuations (of order $\sim$ percent level) can result in up to $\sim$ order of magnitude luminosity variability, owing to the exponential enhancement of the Wien-tail. This variability is well described by the analytical model of \cite{MummeryBalbus22}. This excellent fit between analytical and numerical theory means that the 
(significantly cheaper to compute) analytic theory can be used to probe variability in astrophysical disc systems, and potentially elucidate the nature of disc turbulence.  

By utilising the numerical stochastic-viscosity hydrodynamic models of \cite{Turner23}, we have shown that both the physics of the disc turbulence (in particular the turbulent coherence length), and relativistic viewing angle effects, can have pronounced effects on the observed variability. Increasing the turbulent coherence length (likely set by the disc thickness) results in more pronounced large scale structures to develop in the flow (Fig. \ref{fig:disc_phys}), which results in an increase in luminosity variability. Observing a relativistic disc system at a larger inclination results, {\it at fixed X-ray luminosity}
, in an increase in the observed variability of the disc. This is a result of the ``zooming in'' onto smaller scale structure in the small disc region which is Doppler boosted towards the observer, which dominates the total flux.   As far as the authors are aware this is the first such analysis of relativistic viewing effects. At fixed inclinations we have demonstrated that fainter disc systems have a larger amplitude of variability \citep[Fig. \ref{fig:fit_to_lc}; see also][]{MummeryBalbus22}.

Finally, we argue that tidal disruption events may represent the ideal astrophysical systems with which to observationally probe the small length scale turbulent physics of black hole accretion. Tidal disruption event discs have the fortunate properties of (i) being typically much cooler than observing bandpasses, putting their emission in the Wien tail; (ii) likely having thicker discs (and larger turbulence coherence lengths) than soft-state X-ray binaries; (iii) lacking complicating power-law components in their X-ray spectra, allowing a cleaner look into the inner disc, and (iv) naturally evolving on $\sim$ hour-to-day timescales, owing to their large black hole masses.  Indeed, the tidal disruption event AT2019azh \citep{Hinkle21, Goodwin22} may well already have been observed to show properties predicted by the theory of propagating fluctuations.

\section*{Acknowledgments} 
AM was supported by a Leverhulme Trust International Professorship grant [number LIP-202-014]. SGDT acknowledges support under STFC Grant ST/X001113/1, and previously under an STFC PhD Studentship. For the purpose of Open Access, the authors have applied a CC BY public copyright licence to any Author Accepted Manuscript version arising from this submission. The hydrodynamical simulations used in this work (which were originally presented in \citealt{Turner23}), were performed using resources provided by the Cambridge Service for Data Driven Discovery (CSD3) operated by the University of Cambridge Research Computing Service (\url{www.csd3.cam.ac.uk}), provided by Dell EMC and Intel using Tier-2 funding from the Engineering and Physical Sciences Research Council (capital grant EP/T022159/1), and DiRAC funding from the Science and Technology Facilities Council (\url{www.dirac.ac.uk}). 

\section*{Data availability }
The observational data used in producing this manuscript (Fig. \ref{fig:real-tdes}) is publicly available \citep{Hinkle21}. The numerical data will be shared upon reasonable request with the corresponding author. 

\bibliographystyle{mnras}
\bibliography{andy}

\label{lastpage}
\end{document}